\documentclass[twocolumn,reprint,amsmath,amssymb,aps,superscriptaddress]{revtex4-2}

\usepackage[utf8]{inputenc}
\usepackage{graphicx}
\usepackage{comment}
\usepackage{lipsum}
\usepackage{natbib}

\usepackage[]{hyperref}

\hypersetup{
    colorlinks=true,
    allcolors=blue
}

\makeatletter
\newcommand*{\balancecolsandclearpage}{%
  \close@column@grid
  \clearpage
  \twocolumngrid
}
\makeatother

\let\epsilon\varepsilon

\newcommand*\diff{\mathop{}\!\mathrm{d}}

\newcommand{\sub}[1]{_{\textrm{#1}}}

\newcommand{\figref}[1]{\mbox{Fig.\hspace{0.25em}\ref{#1}}}

\newcommand{\Supp}{supporting material}

\begin{document}

\title{A phase-field model for viscoelastic compressible tumor growth}

\author{Luise Zieger}
\affiliation{Faculty of Mathematics and Computer Science, Technische Universität Bergakademie Freiberg, 09599 Freiberg, Germany}
\author{Min Wu}
\affiliation{Department of Mathematical Sciences, Worcester Polytechnic Institute, Worcester, MA 10605 USA}
\author{Chaozhen Wei}
\affiliation{School of Mathematical Sciences, University of Electronic Science and Technology of China, Chengdu, Sichuan 611731, China}
\author{John Lowengrub}
\affiliation{Department of Mathematics, University of California Irvine, Irvine, CA 92697, USA}
\author{Sebastian Aland}
\affiliation{Faculty of Mathematics and Computer Science, Technische Universität Bergakademie Freiberg, 09599 Freiberg, Germany}
\affiliation{Hochschule für Technik und Wirtschaft Dresden, 01069 Dresden, Germany}
\affiliation{Center for Systems Biology Dresden, 01307 Dresden, Germany}
\affiliation{Cluster of Excellence Physics of Life, TU Dresden, 01062 Dresden, Germany}

\date{\today}

\begin{abstract}
It is well known that growing tumors generate and respond to stress in their local microenvironment. Tissue re-arrangements can relax these mechanical stresses and make the tissue more fluid-like. Further, intricate coupling between mechanotransduction and biochemical signaling leads to complex patterns of growth. To predict the outcomes of these nonlinear interactions, we develop a phase-field model to simulate tumors growing into a surrounding medium taking into account their elastic and viscous properties as well as their compressibilities. We couple continuum modeling of the viscoelastic mechanics to the concentration of a diffusible growth-promoting nutrient in a mass conservative way. The phase-field method is a stable and flexible way to describe the dynamics of arbitrarily shaped tumors. We demonstrate convergence of the phase-field model to a sharp interface model in radially symmetric geometries and can observe progression to stationary tumors. However, our results show that these stationary symmetric tumors are subject to symmetry-breaking instabilities in 2D and 3D driven by two primary mechanisms: (i) elastic buckling instabiliies due to differential growth induced by the nutrient gradient and (ii) instabilities generated by apoptosis-related volumetric loss. Further, tissue fluidity and compressibility can lead to changes in tumor topologies. Our modeling framework provides a robust methodology for investigating how tissue mechanics and growth factor signaling influence the progression and invasive potential of solid tumors. 
\end{abstract}

\maketitle


\section*{Introduction}

The growth of solid tumors is a complex, multiscale process in which mechanical, chemical, and biological factors are intimately coupled \cite{Nia2020,Xu2024}.
When cells proliferate or undergo apoptosis within developing tissues or tumors, mechanical stresses inevitably arise, accumulate, and
deform the surrounding tissue \cite{LeGoff2016,Linke2024,angeli2025}. 
Through the process of mechanotransduction, these stresses feed back to alter fundamental cell behaviors including proliferation, death, differentiation, and
motility \cite{Hsieh2005,Clause2010,Schwartz2018,Chen2023}. 
Simultaneously, biochemical signaling---mediated by diffusible growth factors, cytokines, and cell--cell communication networks---exerts an
independent but intertwined influence on these same cell-level processes \cite{Armingol2021}. 
Tissue rearrangements, driven by cell migration, intercalation, and division, can relieve accumulated mechanical stress \cite{LeGoff2016,David2014}, and cells can dynamically remodel the stiffness of their local microenvironment through the secretion of extracellular matrix (ECM) molecules and matrix-degrading enzymes \cite{Vasudevan2023}. 

A growing body of experimental evidence has established that the mechanical microenvironment plays a critical role in regulating tumor progression \cite{Jain2014,Linke2024,angeli2025}. 
Studies of tumor spheroids embedded in three-dimensional hydrogels have demonstrated that spheroid size, morphology, and internal cell density are strongly sensitive to the stiffness of the surrounding matrix \cite{Helmlinger1997,Cheng2009,mahajan}. 
In stiff hydrogels, high levels of compressive stress can build up around spheroids, which can enable cells to stiffen \cite{Taubenberger2019}, increase cell densities and inhibit tumor growth  \cite{Helmlinger1997}, suppress proliferation and induce apoptosis \cite{Cheng2009}, stimulate local invasion \cite{Tse2012} and resist drug treatment \cite{Rizzuti2020}.
These diverse observations underscore the need for theoretical frameworks capable of accurately coupling tissue mechanics, growth, and biochemical signaling to explain and predict emergent behaviors and complex growth patterns.

Mathematical models of stress in solid tissue \cite{ambrosi2009cell,goriely2017,Xuetal2023,Walker2023,Holzapfel2025,Carotenuto2025} include poroelastic models, which treat the tissue as a poroelastic solid and capture the coupling between solid stress and interstitial fluid flow, morphoelastic models, which are generally used to capture the interaction between mechanical forces and cell proliferation and death, models that combine the two approaches \cite{Armstrong2016}, and new, multiscale continuum models that involve directly upscaling discrete cell-based models \cite{Weady2024}.

Elastic stresses can dissipate due to internal cell processes, such as turnover of structural and adhesion molecules, as well as cell-rearrangements and oriented cell divisions \cite{LeGoff2016}. 
These relaxation mechanisms can occur on time scales that range from minutes to hours \cite{Cox2011,Doubrovinski2017}. This motivates the development of viscoelastic models of tumor mechanics, e.g.,   \cite{Bresch2009,Preziosi2010,Ambrosi2012,lowengrub2021ConfinedTumor,wei2023eulerian,Olaranont2024,garcke2022viscoelastic,Garcke2024,Chockalingam2024,Firouzi2025}.

In this work, we construct, validate, and explore a new model for tumor growth based on a phase-field (diffuse-interface) description of the tumor.
In this approach, the tumor is modeled via a dynamically evolving, smoothed approximation of a characteristic function and the tumor-host boundary has a small but finite thickness, e.g., \cite{provatas2011phase}. 
The phase-field approach is a stable and flexible way to simulate tumor growth without geometric constraints on its shape. This approach also enables large deformations and topological changes to be captured. 

In contrast to previous models that focused on incompressible growth (e.g., \cite{garcke2022viscoelastic,Garcke2024}), we develop a model in which the tumor and surrounding medium are \emph{compressible}. 
This enables the tumor expansion to freely emerge from the interplay of mechanical forces and biochemical signaling, rather than being prescribed as in incompressible models. 
Depending on the choice of mechanical parameters, both compressible and incompressible growth regimes can be recovered. 
To account for tissue rearrangement, we incorporate a Maxwell-type relaxation of shear stresses, leading to an intricate coupling between mechanics and shape dynamics. 
In our framework, the tissue rearrangement is mass conservative.

The overall model is validated using a radially symmetric sharp-interface model \cite{Olaranont2024}. 
Two and three dimensional simulations demonstrate the development of symmetry-breaking instabilities and topological transitions driven by compressive forces, differential growth, tissue fluidity and compressibility. 
Thus, our framework provides a quantitative foundation to elucidate the mechanisms by which physical forces and biochemical signaling networks dictate solid tumor progression.


\section*{Mathematical model}

In this section, we first derive the sharp interface description for a nonlinear elastic, compressible tissue embedded in a surrounding medium, where the tissue growth is coupled with the reaction-diffusion process of growth factors (e.g., nutrients). In the following section, we formulate an equivalent system within a domain characterized by a diffuse interface.

\subsection*{Sharp interface model}\label{sec:sharp_interface_model}

\begin{figure*}[hbt!]
    \centering
    \includegraphics[width=0.85\textwidth]{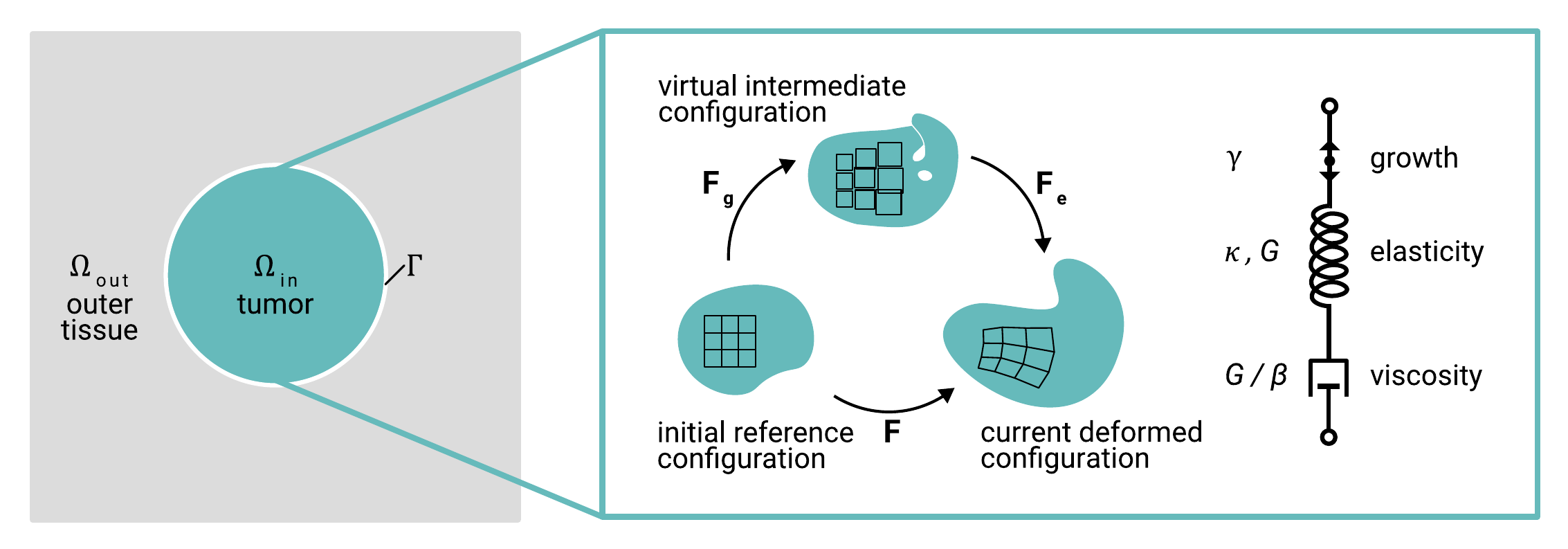}
    \caption{The whole domain is divided into two subdomains $\Omega_\text{in}$ (tumor) and $\Omega_\text{out}$ (host tissue) separated by $\Gamma$. Both tissues are modeled as compressible neo-Hookean elastic materials. 
    Within the tumor domain, growth is represented by a local inelastic tensor ($\mathbf{F}_{\rm g}$) that maps to a virtual, stress-free configuration; an elastic deformation ($\mathbf{F}_{\rm e}$) is then required to ensure geometric compatibility.
    The rearrangement rate $\beta$ accounts for the tissue rearranging activities, resulting in a Maxwell-type stress relaxation. The tumor exhibits isotropic growth with rate $\gamma$ in response to spatial stimuli.}
    \label{fig:sharp_interface_domain}
\end{figure*}

We first consider the dynamic system for tumor growth based on a sharp interface model. Let $\Omega$ be the entire domain, composed of two subdomains, $\Omega_\text{in}$ and $\Omega_\text{out}$, which represent the tumor and the surrounding tissue, respectively. The interface between both subdomains is denoted by $\Gamma$, as shown in \figref{fig:sharp_interface_domain}. We will derive the governing equations for the mechanics of the growing tumor and the dynamics of the interface. 

\subsubsection*{Derivation of the nonlinear elastic tumor growth model}

The theory of morphoelasticity \cite{Rodriguez1994} offers a natural framework to study tissue growth, where the most crucial aspect is accurately describing the connection between growth and elasticity. The fundamental concept here is that during a growth process, the deformation of a body can result from both its active growth and elastic response. Accordingly, the deformation gradient tensor can be decomposed into a growth tensor that captures how the volume is locally added or removed at different parts of the body and an elastic deformation tensor which accounts for the elastic rearrangement needed to maintain the continuum integrity and the elastic response to external loading. The full deformation gradient $\mathbf{F}$ is then multiplicatively decomposed by
\begin{equation*}
    \mathbf{F} = \mathbf{F_e} \mathbf{F_g}.
\end{equation*}

The growth stretch tensor $\mathbf{F_g}$ maps from the initial stress free configuration to a virtual intermediate configuration accounting for volume changes arising from processes such as cell proliferation, apoptosis, or active contractility. The elastic deformation tensor $\mathbf{F_e}$ maps from the intermediate configuration to the current deformed configuration, describing the elastic deformation of the material. 
Physically, $\mathbf{F_g}$ describes the stress-free volumetric changes of local continua due to active growth, which may lead to an incompatible virtual intermediate configuration where the material can overlap or have holes (Fig. \ref{fig:sharp_interface_domain}). Consequently, an additional elastic deformation, $\mathbf{F_e}$, is required to reorganize the material's shape and size, maintaining its integrity. This reorganization generates residual stresses in the current deformed configuration.

Let us consider the change in mass density of the tissue during growth and elastic deformation. Denoting the volumetric variation due to the total geometric deformation, active growth, and elastic deformation as $J = \det(\mathbf{F})$, $J_g = \det(\mathbf{F_g})$, and $J_e = \det(\mathbf{F_e})$, respectively, they satisfy the relation $J=J_eJ_g$. Considering an infinitesimal volume element $V_0$ in the initial configuration with a stress-free initial density $\rho_0$, the density remains constant while the volume changes to $J_gV_0$ during the stress-free active growth. As a result, the mass of the volume element in the intermediate configuration changes from $\rho_0V_0$ to $\rho_0J_gV_0$. Following elastic deformation, the density becomes $\rho$, and the volume changes to $JV_0$, such that the mass of the element in the current deformed configuration is $\rho J V_0=\rho J_g J_eV_0$. Since elastic deformation does not introduce additional mass, mass conservation between the intermediate and current configuration gives $\rho_0J_gV_0=\rho J_g J_e V_0$, which yields the condition
\begin{align}
   \rho J_e=\rho_0=\text{const}.
   \label{eq:mass cons}
\end{align} 

Assuming the dynamics of the growth stretch tensor as in \cite{Olaranont2024},
\begin{equation}
    \frac{d\mathbf{\mathbf{F}_g}}{d t} = \tilde{\boldsymbol{\tau}}\mathbf{\mathbf{F}_g},\qquad  \tilde{\boldsymbol{\tau}}=\frac{\gamma}{d}\mathbf{I}+\tilde{\boldsymbol{\tau}}_D, \label{eq:Fg_evolution}
\end{equation}
where $d/dt = \partial_t+\mathbf{v}\cdot\nabla$ is the material derivative and
we have decomposed the growth rate tensor $\tilde{\boldsymbol{\tau}}$ into its isotropic part with the volumetric growth rate $\gamma=\text{tr}(\tilde{\boldsymbol{\tau}})$ and the deviatoric traceless part $\tilde{\boldsymbol{\tau}}_D$ (i.e., $\text{tr}(\tilde{\boldsymbol{\tau}}_D)=0$). Here, $d$ denotes the number of spatial dimensions. Following \cite{Olaranont2024}, we assume that $\tilde{\boldsymbol{\tau}}_D$ accounts for the isochoric rearrangement of tissue that is both mass-conserving and energy-dissipative and has the following form
\begin{equation}
    \tilde{\boldsymbol{\tau}}_D=\beta\mathbf{\mathbf{F_e}}^{-1}\left(\mathbf{B_e}  - \frac{\text{tr}(\mathbf{B_e})}{d} \mathbf{I}\right)\mathbf{\mathbf{F_e}},
\end{equation}
where $\mathbf{B_e} := \mathbf{F_e}\mathbf{F_e}^\top$ is the left Cauchy-Green tensor and $\beta$ represents the rearrangement rate. Using the identity $\frac{d}{dt}(\det \boldsymbol{A}) = \det \boldsymbol{A} \text{tr} \left(\boldsymbol{A}^{-1} \frac{d \boldsymbol{A}}{dt}\right)$, which holds for any matrix $\boldsymbol{A}$, the volumetric change due to active growth is given by
\begin{equation}\label{eq:Jg_evol}
    \frac{dJ_g}{dt}=\gamma J_g.
\end{equation}

Furthermore, the dynamics of the deformation gradient tensor $d\mathbf{\mathbf{F}}/dt=\nabla\textbf{v}\mathbf{\mathbf{F}}$ gives the total volumetric change rate 
\begin{equation}\label{eq:J_evol}
    \frac{dJ}{dt}=\nabla \cdot \mathbf{v} J.
\end{equation}

Combining the relation \eqref{eq:mass cons} and the governing equations \eqref{eq:Jg_evol} and \eqref{eq:J_evol}, we can obtain the mass balance equation
    \begin{align}
    \frac{d \rho}{d t} &=  \rho(\gamma - \nabla \cdot \mathbf{v}). 
    \label{eq:mass cons2}
\end{align}

Moreover, the dynamics of the elastic deformation tensor in presence of growth and tissue rearrangement is then given by 
\begin{equation}
    \frac{d\mathbf{\mathbf{F_e}}}{d t} = \left(\nabla \mathbf{v} - \frac{\gamma}{d}\mathbf{I}\right)\mathbf{F_e} - \beta \left(\mathbf{B_e}  - \frac{\text{tr}(\mathbf{B_e})}{d} \mathbf{I}\right) \mathbf{F_e}, \label{eq:Fe_evol}
\end{equation}
which yields the dynamics of $J_e$, 
\begin{equation}
    \frac{d J_e}{d t} = J_e \left(\nabla \cdot \mathbf{v} - \gamma\right), \label{eq:Je_evol}
\end{equation}
that is consistent with the mass balance \eqref{eq:mass cons}.

By \eqref{eq:Fe_evol}, we can obtain the frame-invariant evolution of the left Cauchy-Green tensor $\mathbf{B_e}$ \cite{Olaranont2024}
\begin{align*}
    \frac{d}{d t} \mathbf{B_e} &= \frac{d \mathbf{F_e}}{d t} \mathbf{F_e}^\top + \mathbf{F_e} \frac{d \mathbf{F_e}^\top}{d t} \\
    &= \nabla \mathbf{v} \mathbf{B_e} + \mathbf{B_e} \nabla \mathbf{v}^\top - \frac{2\gamma}{d}\mathbf{B_e}- 2\beta \mathbf{B_e}\left(\mathbf{B_e}  - \frac{\text{tr}(\mathbf{B_e})}{d} \mathbf{I}\right)
\end{align*}
or equivalently 
\begin{equation}
    \overset{\kern-0.5em\triangledown}{\mathbf{B_e}} = - \frac{2}{d} \gamma \mathbf{B_e} - 2 \beta \mathbf{B_e} \left( \mathbf{B_e}  - \frac{\text{tr}(\mathbf{B_e})}{d} \mathbf{I}\right), 
\end{equation}
where the notation $\overset{\kern0.25em\triangledown}{\mathbf{q}}=d\mathbf{q}/dt-\nabla\mathbf{v}\mathbf{q}-\mathbf{q}\nabla\mathbf{v}^T$ denotes the upper convected derivative of a tensor $\mathbf{q}$.

\subsubsection*{Compressible neo-Hookean elasticity}

Soft tissues can be considered as isotropic, compressible neo-Hookean elastic materials \cite{wei2023eulerian}. 
Decomposing the corresponding Cauchy stress into volumetric and deviatoric components yields:
\begin{equation*}
    \mathbf{S} = \kappa (J_e - 1)\mathbf{I} + G J_e^{-\frac{2+d}{d}}\left(\mathbf{B_e}-\frac{1}{d}\text{tr}(\mathbf{B_e})\mathbf{I}\right),
\end{equation*}
where $\kappa$ and $G$ denote the bulk and shear moduli, respectively. 

Assuming over-damped dynamics where the frictional force is proportional to the tissue velocity, the balance of the elastic force and the frictional force provides 
\begin{equation}
    \zeta\mathbf{v} = \nabla \cdot \mathbf{S} ~\qquad\mbox{in}~\Omega_\text{in}, \label{eq: velocity}
\end{equation}
where $\zeta$ denotes the friction coefficient. The equation can be equipped with various boundary conditions. To describe a single tissue in $\Omega_\text{in}$ without confinement, the condition $\mathbf{S}\cdot\mathbf{n}=0$ is imposed on $\Gamma$.
If a confining outer tissue is present in $\Omega_\text{out}$, the definitions of stress and strain need to be extended into this domain, e.g.  
\begin{align*}
    \mathbf{S}_i &= \kappa_i (J_e - 1)\mathbf{I} + G_i J_e^{-\frac{2+d}{d}}\left(\mathbf{B_e}-\frac{1}{d}\text{tr}(\mathbf{B_e})\mathbf{I}\right), \\ &~\qquad\mbox{in}~\Omega_i, ~i\in\{{\rm in, out}\},
\end{align*}
with $\kappa\sub{in}, G\sub{in}$ and $\kappa\sub{out}, G\sub{out}$ the bulk and shear moduli of the inner and outer tissue, respectively. 
The natural boundary condition in absence of surface forces is $\left[\mathbf{S}\right]_{\rm out}^{\rm in}\cdot\mathbf{n}=0$ on $\Gamma$, where $[\cdot]_{\rm out}^{\rm in}$ denotes the jump across the interface $\Gamma$. 		
In this case, also the velocity equation \eqref{eq: velocity} can be similarly extended into both domains, coupled by the continuity condition $\left[\mathbf{v}\right]_{\rm out}^{\rm in}=0$ on $\Gamma$.
The phase-field approach presented below will alleviate the solution of the complex coupled system as it naturally regularizes the velocity and stresses across the interface.

\subsubsection*{Nutrient}

The volumetric growth rate $\gamma$ is naturally coupled with the field of a diffusible growth-promoting factor. Assuming that the nutrient is supplied through the vasculature of the outer tissue and undergoes diffusion and uptake within the tumor, the nutrient concentration $c$ is governed by a reaction-diffusion equation with Dirichlet boundary condition. Furthermore, we assume that the characteristic time of this reaction-diffusion process is much smaller then the time scale of growth $1/\gamma$, which leads to the quasi-stationary reaction-diffusion equation
\begin{align}
\begin{array}{r l}
     - L^2 \Delta c + c = 0 &  \text{ in } \Omega_\text{in},\\
     c = 1 &  \text{ in } \Gamma \cup \Omega_\text{out},
\end{array} \label{eq:nutrient_concentration}
\end{align}
where $L^2$ is the squared diffusion length (i.e. diffusion constant times turn over time of nutrient). The Dirichlet boundary condition assumes a saturation value $c=1$ outside the tumor and the nutrient diffuses from the boundary to the tumor central region. 

Then the volumetric growth rate is coupled with the local concentration of nutrient by
\begin{equation*}
    \gamma(c) = \lambda\sub{p} c - \lambda\sub{a},
\end{equation*}
where $\lambda\sub{p} c$ and $\lambda\sub{a}$ represent the local rates of cell proliferation and apoptosis, respectively. Thus, $\gamma$ decreases from the boundary to the center of the tumor due to decreasing nutrient concentration.

\subsection*{Phase-field approach} \label{sec:phase_field_formulation}

\begin{figure}[ht]
    \centering
    \includegraphics[width=\linewidth]{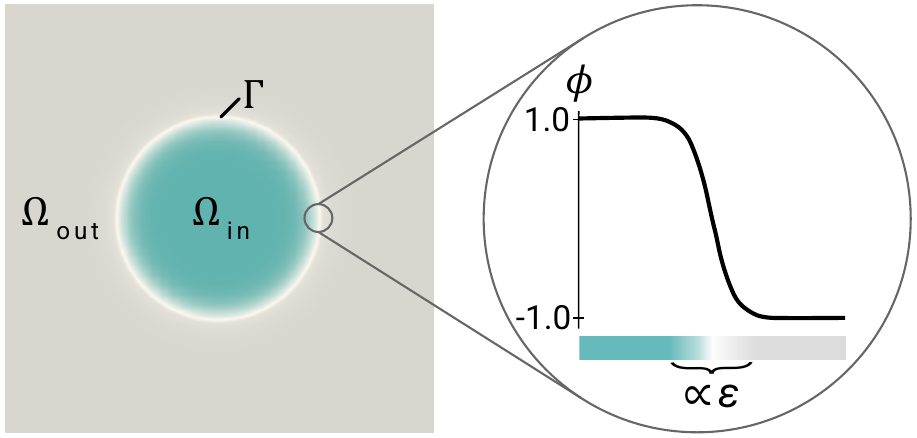}
    \caption{The phase field representation of a circular tumor in a rectangular domain with smooth transition at the interface and an interface thickness proportional to $\epsilon$.}
    \label{fig:phase-field}
\end{figure}

The phase-field approach is a flexible way to track the boundary of moving domains, including large deformations, topological changes and (wall) contact. In the following, we introduce a phase-field formulation of the above system of equations, coupling the mechanics within the tumor with the mechanics of the surrounding tissue.  Let $\phi$ be a phase-field variable that distinguishes between the tumor domain ($\phi \approx 1$) and the outer domain ($\phi \approx -1)$ within a computational domain $\Omega$. Thus, $\Omega_\text{in}$ and $\Omega_\text{out}$ are approximated by the domains where $\phi \approx 1$ and $\phi \approx -1$, respectively, with a smooth transition between the two phases. The interface can be associated with an intermediate level set ($\phi = 0$) of the phase-field function, see Fig.~\ref{fig:phase-field}. 

\noindent Initially, we prescribe the phase field by 
\begin{equation*}
    \phi = \tanh\left(r/\left(\sqrt{2}\epsilon\right)\right),
\end{equation*}
where $r$ denotes the signed distance to $\Gamma$ and is positive inside $\Omega_\text{in}$. In order to track changes in the tumor shape, the phase field will be advected with the velocity $\mathbf{v}$. This is described by the convective Cahn-Hilliard equation 
\begin{align}
    \partial_t \phi + \mathbf{v} \cdot \nabla \phi &= \nabla \cdot (M \nabla \mu), \label{eq:cahn-hilliard}\\
    \mu &= f'(\phi) - \epsilon^2 \Delta \phi, \label{eq: mu}
\end{align}
in the domain $\Omega$. Here, $\mathbf{v}$,  $\mu$ and $M$ are the velocity, chemical potential and mobility, respectively. The parameter $\epsilon > 0$ scales with the thickness of the interface region. The double well potential $f$ is given by $f(\phi) = 1 / 4(\phi^2 - 1)^2$, which leads to the values $\phi \approx \pm1$ in the two distinct phases (Fig.~\ref{fig:phase-field}).

We couple the previously described tissue mechanics to the phase-field variable by defining the elastic moduli to be phase-dependent,
\begin{align*}
    \kappa(\phi) &= \kappa\sub{in}\frac{\tilde\phi+ 1}{2}  + \kappa\sub{out} \frac{1 - \tilde\phi}{2}, \\
    G(\phi) &=  G\sub{in}\frac{\tilde\phi+ 1}{2}  + G\sub{out} \frac{1 - \tilde\phi}{2},
\end{align*}
where $\tilde\phi$ is a version of $\phi$ truncated to $[-1,1]$ to ensure that the parameters lie in between the values of the pure phases. 
The diffuse form of the stress equation is now
\begin{equation}
    \mathbf{S} = \kappa(\phi) (J_e - 1)\mathbf{I} + G(\phi) J_e^{-\frac{2+d}{d}}\left(\mathbf{B_e}-\frac{1}{d}\text{tr}(\mathbf{B_e})\mathbf{I}\right). \label{eq:stress}
\end{equation}

Analogously, the relaxation-rate may be different in the two tissues, as modeled by
\begin{align} \label{eq: beta of phi}
    \beta(\phi) &= \beta\sub{in}\frac{\tilde\phi+ 1}{2}  + \beta\sub{out} \frac{1 - \tilde\phi}{2}.
\end{align}
The evolution of the left Cauchy-Green tensor $\mathbf{B_e}$ becomes
\begin{equation} \label{eq: Be}
    \overset{\kern-0.5em\triangledown}{\mathbf{B_e}} = - \frac{2}{d} \gamma(\phi,c) \mathbf{B_e} - 2 \beta(\phi) \mathbf{B_e} \left( \mathbf{B_e}  - \frac{\text{tr}(\mathbf{B_e})}{d} \mathbf{I}\right), 
\end{equation}
which is phase-dependent. 
We assume that growth occurs exclusively within the tumor by defining
\begin{equation} \label{eq: gamma}
    \gamma(\phi,c) =  \left( \lambda\sub{p} c - \lambda\sub{a}\right) \frac{\phi + 1}{2},
\end{equation}
resulting in a growth rate of $\gamma(\phi) = 0$ in the outer medium, where $\phi = -1$. Rather than reformulating Eq. (\ref{eq:nutrient_concentration}) for the nutrient concentration $c$ using a phase-field approach \cite{LiLowengrubRätzVoigt}, the embedded boundary method (e.g., \cite{johansen1998,balajewicz2014reduction}) is used instead, which is more accurate for larger values of the interface thickness $\epsilon$.

The system of equations is closed by the force balance for the velocity field. Assuming equal friction in both tissues gives the equation 
\begin{equation}
    \zeta \mathbf{v} = \nabla \cdot \mathbf{S} ~\qquad\mbox{in}~\Omega, \label{eq: v pf}
\end{equation}
as before.

\subsection*{Remark on thermodynamical consistency}

We note that it is also possible to construct the diffuse-interface model, based on an energy-dissipative modeling approach, as shown in the following. 

For a compressible neo-Hookean material, the strain energy density function (per unit stress-free volume) in $d$ dimensions is given by
\begin{equation*}
    W=\frac{\kappa(\phi)}{2}(J_e -1)^2 + \frac{G(\phi)}{2}\left(J_e^{-\frac{2}{d}}\mathrm{tr}(\mathbf{B_e})-d\right).
\end{equation*}

Including the Ginzburg-Landau energy of the diffuse interface, leads to the free energy of the system
\begin{equation*}
    E = \int_\Omega \left(\frac{\sigma\epsilon}{2} |\nabla\phi|^2+\frac{\sigma}{4\epsilon}\left(\phi^2-1\right)^2\right)\mathrm{d}x + \int_\Omega J_e^{-1}W \mathrm{d}x,
\end{equation*}
where $\sigma$ denotes the surface tension parameter. 
Taking the time derivative of $E$ in absence of growth (i.e., $\gamma=0$) yields
\begin{align*}
    \frac{\diff E}{\diff t} =& \int_\Omega \left( \frac{\sigma}{\epsilon}\mu\partial_t\phi + \mathbf{S}:\nabla\mathbf{v}+J_e^{-1}\frac{\partial W}{\partial\phi}\partial_t \phi\right) \\
    &~~~ - \beta G J_e^{-\frac{2+d}{d}}  \left( \mathbf{B_e}  - \frac{\text{tr}(\mathbf{B_e})}{d} \mathbf{I}\right)^2 \mathrm{d}x, 
\end{align*}
where $\mu$ and $\mathbf{S}$ are defined as in Eqs. \eqref{eq: mu}-\eqref{eq:stress}.
Assuming advection-diffusion kinetics for the phase-field, i.e. 
\begin{align} \label{eq: phi thermo}
    \partial_t\phi + \mathbf{v}\cdot \nabla\phi = -\nabla\cdot \mathbf{j},
\end{align}
the rate of change of the energy becomes
\begin{align*}
    \frac{\diff E}{\diff t} = &\int_\Omega \mathbf{v}\cdot\left(-\nabla\cdot\mathbf{S}  - \frac{\sigma}{\epsilon}\mu\nabla\phi -J_e^{-1}\frac{\partial W}{\partial \phi}\nabla\phi \right)  \\
    &~~~ + \mathbf{j} \cdot  \nabla\left( \frac{\sigma}{\epsilon}\mu+J_e^{-1}\frac{\partial W}{\partial \phi} \right)  \\ 
    &~~~ - \beta G J_e^{-\frac{2+d}{d}}  \left( \mathbf{B_e}  - \frac{\text{tr}(\mathbf{B_e})}{d} \mathbf{I}\right)^2 \mathrm{d}x, 
\end{align*}
where boundary terms were dropped by assuming a no-slip condition on $\partial\Omega$. 
Hence, choosing $\mathbf{j} = -M\nabla\tilde{\mu}$  and 
\begin{align} \label{eq: v thermo}
    \mathbf{v} &:= \frac{1}{\zeta}\left(\nabla\cdot \mathbf{S}  + \frac{\sigma}{\epsilon}\tilde{\mu}\nabla\phi  \right), \\
    \tilde{\mu}&:= \mu + \frac{\epsilon}{\sigma}J_e^{-1}\frac{\partial W}{\partial \phi}, \label{eq: mu tilde}
\end{align}
leads to energy dissipation of the thermodynamically consistent  system \eqref{eq: mu}-\eqref{eq: Be}, \eqref{eq: phi thermo}-\eqref{eq: mu tilde} in absence of active growth. 

When growth is present, the corresponding $\gamma$-term in Eq.~\eqref{eq: Be} introduces an additional contribution to the rate of change of the energy. With the above choice of $\mathbf{j}$ and $\mathbf{v}$, we  obtain 
\begin{align*}
    \frac{\diff E}{\diff t} =& \int_\Omega -{\zeta} |\mathbf{v}|^2 - \frac{\sigma}{M\epsilon} |\mathbf{j}|^2 - \beta G J_e^{-\frac{2+d}{d}}  \left( \mathbf{B_e}  - \frac{\text{tr}(\mathbf{B_e})}{d} \mathbf{I}\right)^2 \mathrm{d}x \\
    +& \gamma\int_\Omega
    -\frac{1}{d}\mathbf{S}:\mathbf{I} + J_e^{-1}W~ 
    \mathrm{d}x,
\end{align*}
where the first line is dissipative and the second line -- scaled with $\gamma$ -- describes the work performed by active growth.

The resulting consistent equations perform well when $\frac{\partial W}{\partial \phi}=0$ (i.e. $\kappa_{\textrm{in}}=\kappa_{\textrm{out}}$ and $G_{\textrm{in}}=G_{\textrm{out}}$). However, in numerical experiments with strong elasticity contrast, we observed that the additional elastic contribution in $\tilde{\mu}$ leads to excessive compression of the tumor and drives the phase field away from the pure-phase values of -1 and 1.
To address this issue, we omit the elastic contribution in the chemical potential and use  $\mu$ instead of $\tilde{\mu}$ throughout this paper. As a result, we recover essentially the same system as before, given by Eqs. \eqref{eq:nutrient_concentration}-\eqref{eq: v pf}, but now including an additional surface tension contribution. As we will show in the next section, this system approximates the corresponding energy-dissipative sharp interface model well as $\varepsilon\rightarrow 0$.


\section*{Validation}

To validate our phase-field model, we compare growth dynamics with a recently developed radially symmetric sharp-interface model of tissue growth and mechanics based on the elasticity theory of compressible tissues and stress relaxation \cite{Olaranont2024}. 
To prevent potential morphological instabilities from confounding the results, we use a radially symmetric variant of our phase-field model, detailed in the \Supp  ~(see Sec. on 1D radially symmetric model). Indeed, 2D and 3D simulations shown later in the Numerical Results section indicate that symmetry-breaking instabilities generically occur.
In the following two subsections, we present the results of validation studies of unconfined and confined growth.

\subsection*{Unconfined growth}

\begin{figure*}[hbt!]
\centering

\includegraphics[width=1\textwidth]{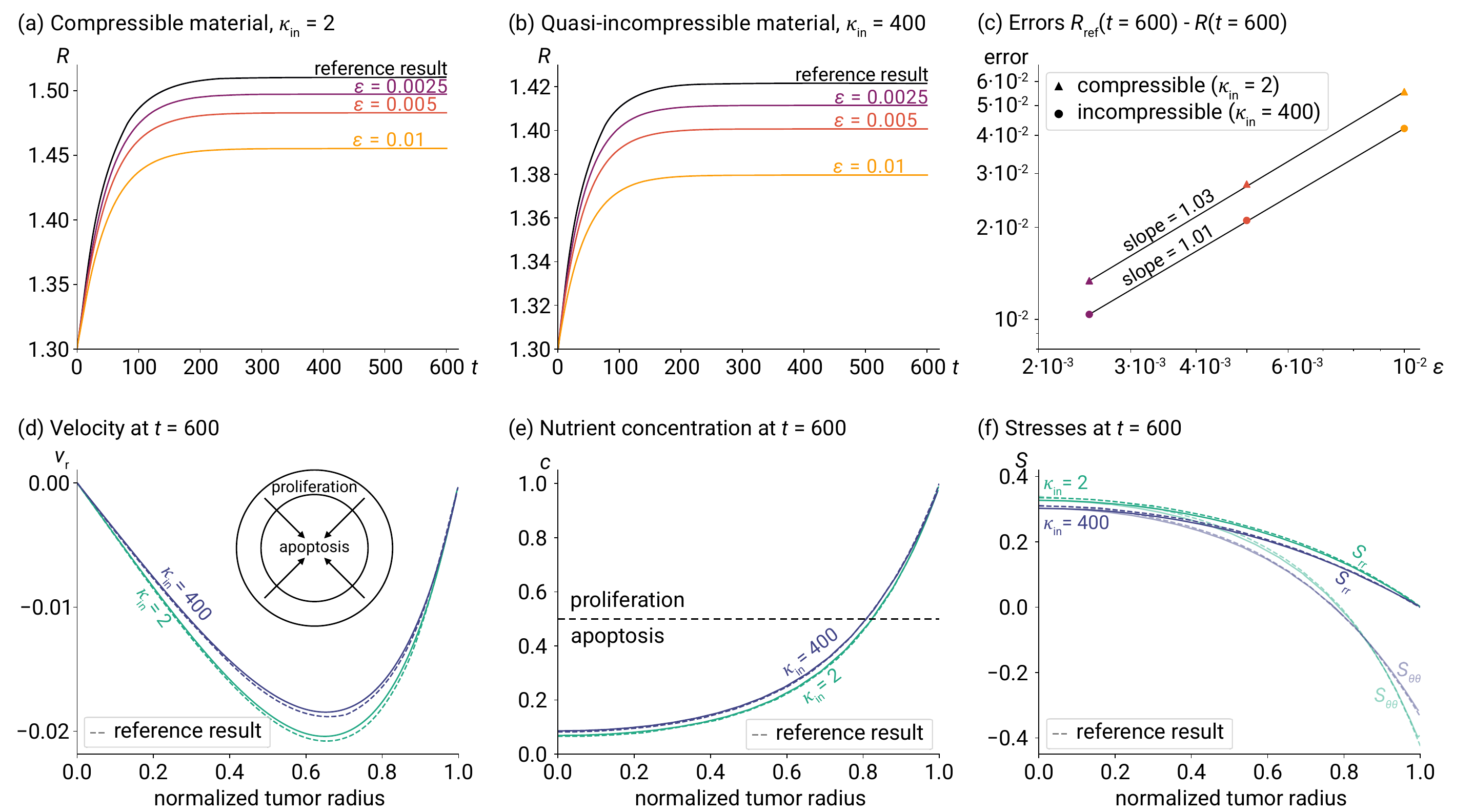} 

\caption{Comparison of the model results with a sharp interface solution without outer medium \cite{Olaranont2024} (unconfined growth). Both models employ radial symmetry, resulting in spherical growth without geometric changes. Top: Evolution of the tumor radius over time for different degrees of interface resolution. The material being (a) compressible, $\kappa\sub{in} = 2, G\sub{in}=1$, or (b) quasi-incompressible, $\kappa\sub{in} = 400, G\sub{in}=1$. (c) The results indicate first order convergence in the interface thickness $\epsilon$ in both examined cases. Bottom: (d) Velocity, (e) nutrient concentration and (f) stresses along the polar axis from the center to the tumor boundary in the stationary state (time point $t=600$). The compressible as well as the quasi-incompressible results match with the reference data (dashed lines). Further simulation parameters: initial tumor radius $R_\Omega= 3, R_0=1.3, \zeta=6.25, M=1.6$, $\beta=0.3$, $\lambda\sub{p}=0.2, \lambda\sub{a}=0.1, L=0.3$.}
\label{fig:ResultMatching_Unconfined}
\end{figure*}

This test case considers a single spherical tumor of initial radius $R_0=1.3$ without a host tissue ($\kappa_\text{out}=G_\text{out}=0$). Because the overall proliferation exceeds apoptosis the tumor grows over time, see Fig.~\ref{fig:ResultMatching_Unconfined}(a,b).
The increase in tumor size progressively reduces the nutrient concentration in the center, slowing down overall growth until the tumor radius converges to a stationary value. In the stationary state, a proliferating rim persists at the tumor boundary, where proliferating material is transported inward toward the tumor center to compensate for local apoptosis, see Fig.~\ref{fig:ResultMatching_Unconfined}(d) and Movie S1 in the \Supp.

A comparison between our model results and the reference data \cite{Olaranont2024} is conducted for two different mechanical parameter regimes of either a compressible ($\kappa\sub{in}/G\sub{in}=2$) or quasi-incompressible  ($\kappa\sub{in}/G\sub{in}=400$) material, see Fig.~\ref{fig:ResultMatching_Unconfined}.
In both cases we observe a clear first order convergence in $\epsilon$ towards the tumor size of the sharp interface reference solution as shown in Fig.~\ref{fig:ResultMatching_Unconfined}(c). The chosen value of $\zeta$ is sufficiently small to keep the system close to mechanical equilibrium ($\nabla \cdot \mathbf{S} = 0$) at all times, thereby ensuring good agreement between the time evolution of the model-predicted radius and the reference solution.
For the smallest interface thickness we find excellent agreement of velocity, concentration and stress fields Fig.~\ref{fig:ResultMatching_Unconfined}(d-f). After cells proliferate at the boundary, they flow inward towards the tumor center because the density decreases as there is more apoptosis than proliferation because the nutrient levels are low. Accordingly,
the radial stresses are positive (tensile). On the other hand, the circumferential stresses are compressive at the tumor boundary and transition to tensile in the tumor center.

\subsection*{Confined growth}

The reference model is also capable of simulating a compressible elastic tumor spheroid enclosed in an incompressible elastic gel \cite{Olaranont2024}. In particular, they model an infinitely large outer medium, assuming a free boundary condition at $r=\infty$, which does neither grow ($\gamma=0$) nor rearrange ($\beta=0$). To approximate this setup with the proposed phase-field model, we choose parameters of a quasi-incompressible outer tissue ($\kappa\sub{out}=400, G\sub{out}=5$) and select a domain size significantly larger than the tumor size. 

As observed previously, the tumor grows over time. As the tumor radius increases, the nutrient concentration diminishes in the center. Eventually, the balance between proliferating and apoptotic cells leads to the convergence of the radius toward a stationary value.

\begin{figure*}[hbt!]
\centering
\includegraphics[width=1\textwidth]{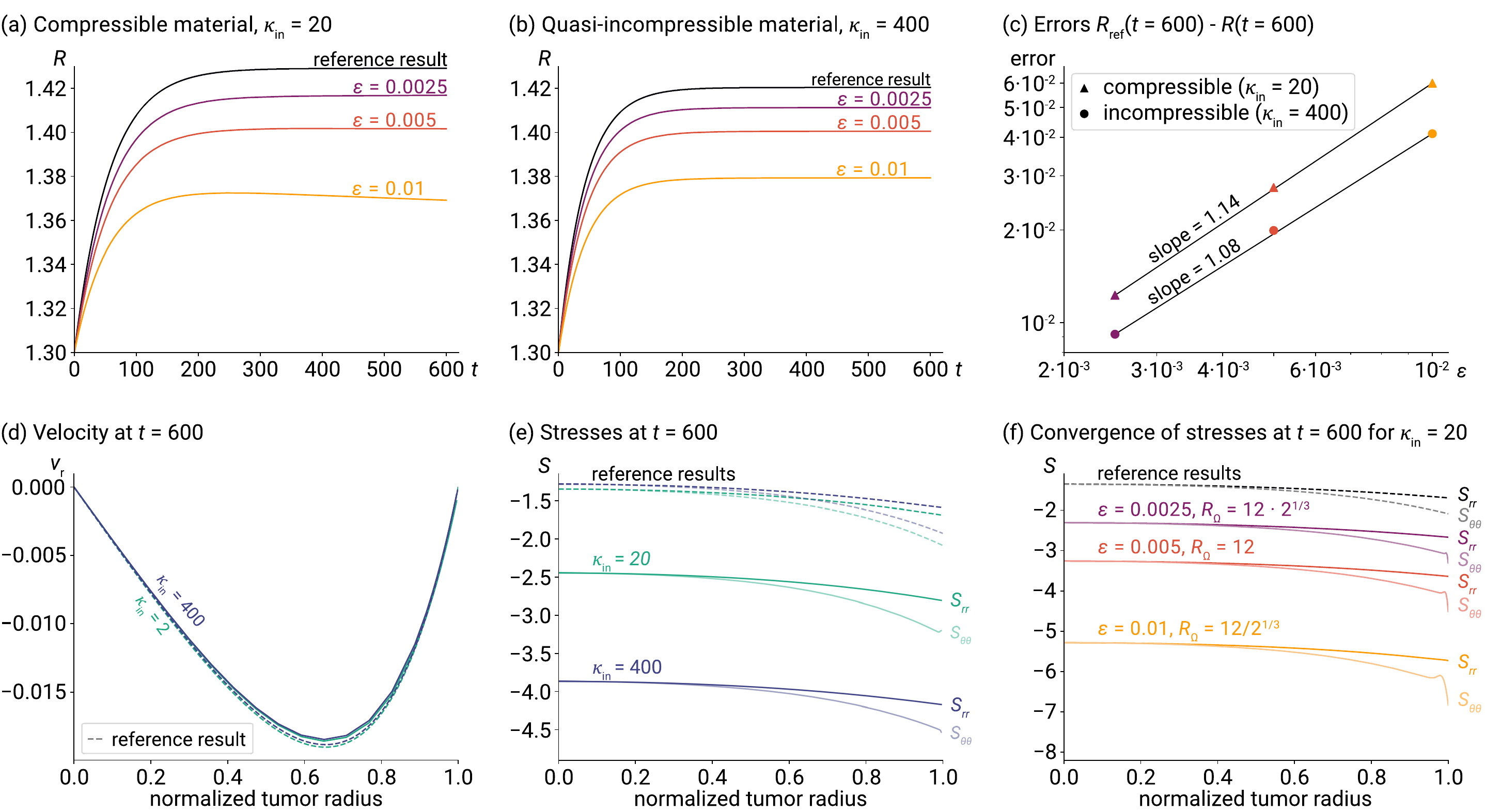} 

\caption{Comparison of the model results with a sharp interface solution including an outer medium \cite{Olaranont2024}. Both models employ radial symmetry, resulting in spherical growth without geometric changes. Top row: The evolution of the tumor radius over time for different degrees of interface resolution. The tumor material being (a) compressible, $\kappa\sub{in}=20, G\sub{in} = 1$, or (b) quasi-incompressible, $\kappa\sub{in}=400, G\sub{in} = 1$, is enclosed by a quasi-incompressible outer material, $\kappa\sub{out}=400, G\sub{out}=5$.  (c) The results provide first order convergence in the interface thickness $\epsilon$ in both examined cases. Bottom rows: (d) Velocity and (e) stresses along the polar axis from the center to the tumor boundary in the stationary state (time point $t=600$) plotted together with the reference data (dashed lines). (f) Influence of interface resolution and domain size on the 
stresses inside the tumor plotted along the polar axis at time $t=600$. The results provide first-order convergence in both $\epsilon$ and $(R_\Omega)^{1/3}$. Further simulation parameters are $R_\Omega= 3, R_0=1.3, \zeta=6.25, M=1.6, \beta\sub{in}=0.3, \beta\sub{out}=0, \lambda\sub{p}=0.2, \lambda\sub{a}=0.1, L=0.3$.}
\label{fig:ResultMatching_Confined}
\end{figure*}

Fig.~\ref{fig:ResultMatching_Confined} shows the comparison between our model results and the reference data conducted for two different mechanical parameter regimes: one with a compressible material ($\kappa\sub{in}/G\sub{in}=20$) and the other with a nearly incompressible material ($\kappa\sub{in}/G\sub{in}=400$).
In both cases we observe a clear first-order convergence in $\epsilon$ towards the tumor size of the sharp interface reference solution, see Fig.~\ref{fig:ResultMatching_Confined}(c). 

For the smallest interface thickness, we observe excellent agreement in the velocity fields, see Fig.~\ref{fig:ResultMatching_Confined}(d). However, discrepancies arise in the stress solutions, where our results perfectly recover the slope of the reference stress but exhibit a constant offset, Fig.~\ref{fig:ResultMatching_Confined}(e). This offset can be attributed to several factors, including the presence of a diffuse interface, the approximation of incompressibility in the outer medium, and the finite domain size. In a finite domain, compressive stresses accumulate and are not released to the surroundings. By refining the interface thickness and increasing the domain size simultaneously, we observe first-order convergence in both the interface thickness $\epsilon$ and $(R_\Omega)^{1/3}$, where $R_\Omega$ denotes the domain radius, towards the stress in the reference solution, see Fig.~\ref{fig:ResultMatching_Confined}(f). Notably, the kinks in the deviatoric stress near the tumor boundary can be attributed to the diffuse interface, and their amplitude decreases when the interface width is refined.

\section*{Numerical Results}
\label{numerical results}

We next study the effect of differential growth in two and three dimensions for various mechanical parameters. We use a finite difference method to perform the simulations, which is described in \Supp.

\subsection*{Instabilities: elastic buckling and invasive fingering}

\begin{figure*}[hbt!]
    \includegraphics[width=\textwidth]{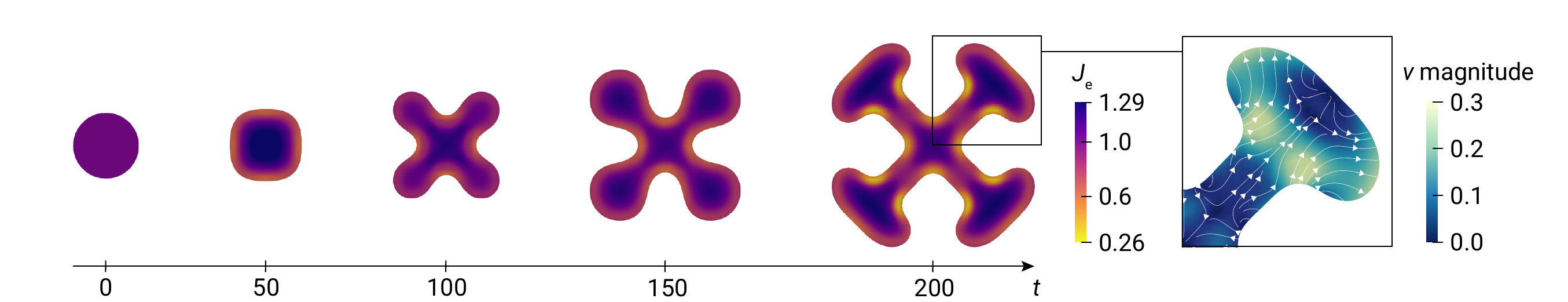}
    \caption{Simulation of the evolution of an initially circular 2D tumor. Snapshots of the tumor shape at the time points $t=0, 50, 100, 150$, and $200$. Small values of $J_e$ at the boundaries indicate compression ($J_e<1$) and high density ($\rho=\rho_0/J_e$). Last panel: direction and magnitude of the material velocity of a tumor section at $t=200$. Simulation parameters: $L_x=L_y=8, h=0.08, R_0=0.8, \epsilon=0.09, \zeta=6.25, M=0.01, \kappa\sub{in} = 2, G\sub{in} = 1, \kappa\sub{out} = G\sub{out} = 0, \beta\sub{in} = \beta\sub{out}=0.05, \lambda\sub{p} = 0.2, \lambda\sub{a} = 0.1, L= 0.3.$ }
    \label{fig:2dEvolution}
\end{figure*}

Starting from a circular tumor, the initial dynamics in higher dimensions resemble those observed in the previous radially symmetric case (Fig.~\ref{fig:2dEvolution}). As before, compressive stress builds up along the tumor periphery as the tumor approaches the stationary state. In 2D and 3D, however, this stress can be partially released through symmetry-breaking deformations that arise from anisotropies introduced by the square computational domain and the Cartesian grid. We observe this behavior in Fig.~\ref{fig:2dEvolution}, where the tumor boundary develops a smoothed square-like shape. This 4-fold symmetry appears as a result of the Cartesian mesh rather than the Cartesian domain, as confirmed in numerical tests (results not shown). Once established, the corner regions experience accelerated proliferation, leading to enhanced outgrowth and the formation of finger-like protrusions. 
In general, we observe material flow from compressed regions ($J_e<1$) towards stretched regions ($J_e>1$).  The tight interplay of these flows with the mechanics of the material lead to complex growth dynamics and shape evolution. 

\begin{figure*}[hbt!]
\centering
\includegraphics[width=\textwidth]{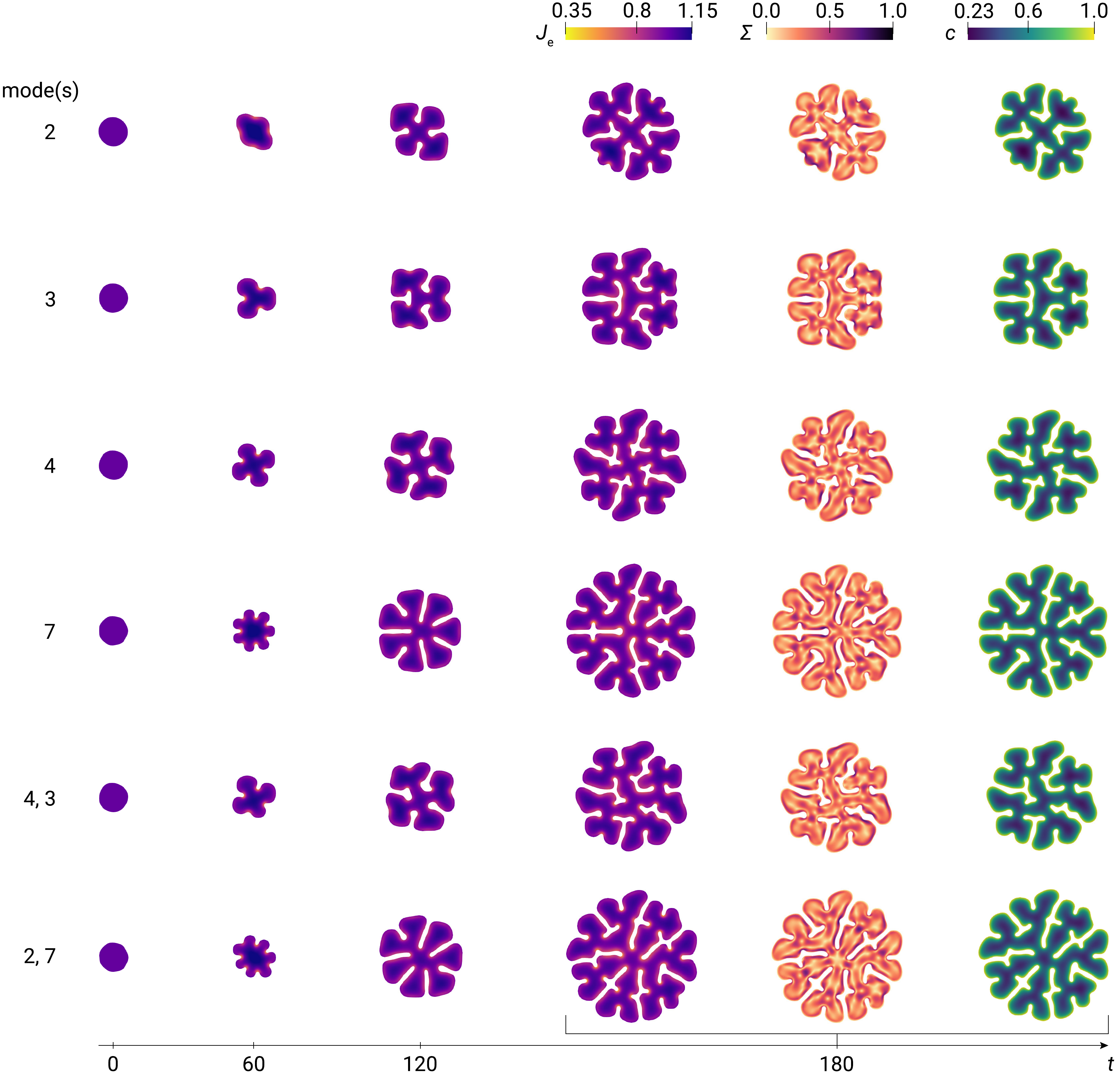}
\caption{Evolution of 2D spherical tumors initialized with sinusoidal perturbations. First four rows: Single mode with $k$ sine peaks ($k=2,3, 4, 7$ from top to bottom). Last two rows: Combinations of even and odd modes. At $t=180$, in addition to the elastic volume variation $J_e$, the stress anisotropy $\Sigma = \lambda_1 - \lambda_2$ (for eigenvalues $\lambda_i$ of $S$), as well as the nutrient concentration field $c$, are shown. Simulation parameters: $L_x=L_y=8, h=0.04, R_0=0.8, \epsilon=0.045, \zeta=6.25, M=0.001, \lambda\sub{p} = 0.2, \lambda\sub{a} = 0.1, L= 0.3, \beta\sub{in} = \beta\sub{out}=0.1, \kappa\sub{in} = 2, G\sub{in} = 1, \kappa\sub{out} = G\sub{out} = 0.$}
\label{fig:perturbed_fingering2}
\end{figure*}

When the initial shapes of the tumor are perturbed explicitly, the number of protrusions correlates to these initial perturbations. This is shown in Fig. \ref{fig:perturbed_fingering2} where the evolutions of 2D tumors with different small sine perturbations are presented. This suggests that all wavenumbers experience positive growth rates, which is consistent with the fact that we have not yet accounted for cell-cell adhesion, e.g., surface tension at the tumor boundary, that stabilizes such dynamics \cite{cristini2003,Pham2018}. See the \Supp ~(Sec. on the influence of surface tension).

Notably, despite the pronounced morphological evolution observed in Fig.~\ref{fig:perturbed_fingering2}, no topological transitions occur.
Instead, the growing and branching protrusions maintain a relatively fixed separation distance. 
This distance is found to scale linearly with $\epsilon$, as we show in the \Supp ~(see Fig. S2). There, we explain in detail how the separation between protrusions is driven by effective repulsion between the interfaces, which arises from shear stress in the diffuse interface region, thereby inhibiting any change in the system's topology. From here on, all presented simulations are performed using the 2, 7 mode.

\begin{figure*}[hbt!] 
    \centering
    \includegraphics[width=\textwidth]{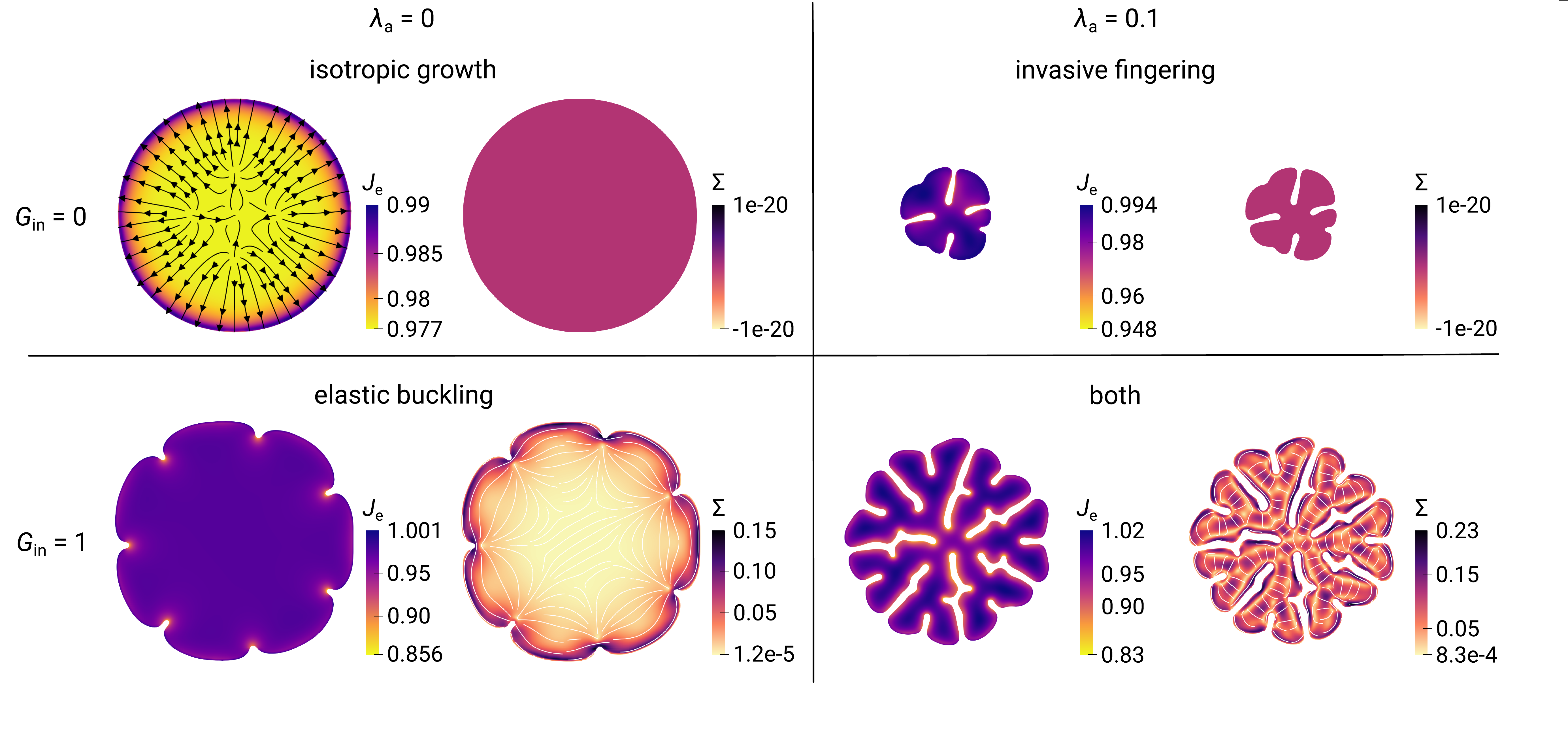}
    \caption{Tumor growth patterns at $t=240$ under varying mechanical and apoptotic conditions. Top row: $G\sub{in}=0$, bottom row: $G\sub{in}=1$; left column: $\lambda\sub{a}=0, \lambda\sub{p}=0.06$, right column: $\lambda\sub{a}=0.1, \lambda\sub{p}=0.2$. In the $\Sigma$ plots, streamlines indicate the direction of the eigenvector of the stress tensor corresponding to the smallest eigenvalue. Further simulation parameters: $L_x=L_y=16, h=0.04, R_0=0.8, \epsilon=0.045, \zeta=6.25, M=0.001, L= 0.3, \beta\sub{in} = \beta\sub{out}=0.3, \kappa\sub{in} = 2, \kappa\sub{out} = G\sub{out} = 0$. 
    }
    \label{fig:instabilities}
\end{figure*}

While the number and geometry of protrusions are influenced by the initial tumor shape, the underlying mechanisms driving these instabilities remain to be clarified. To this end, we next investigate the emergence of different instabilities by systematically varying the mechanical and apoptotic properties of the tissue in our model. Specifically, in the context of differential growth, we explore the effects of shear elasticity, quantified by the shear modulus $G\sub{in}$, and apoptosis, represented via the constant apoptosis rate $\lambda\sub{a}$, on tumor morphology, see Fig.~\ref{fig:instabilities}. By selectively switching these parameters on or off, we observe distinct growth behaviors. 
When both shear elasticity and apoptosis are present
are absent ($G\sub{in}=0,\lambda\sub{a}=0$), tumors expand nearly circularly and isotropically, indicating that the shape instabilities are suppressed, consistent with  \cite{cristini2003,Pham2018}. When apoptosis is suppressed ($\lambda\sub{a}=0$) and shear elasticity is present ($G\sub{in}>0$), tumors develop crease-like boundary patterns, a characteristic feature of elastic buckling induced by differential growth. 
In contrast, when shear elasticity is removed ($G\sub{in}=0$) while apoptosis remains active ($\lambda\sub{a}>0$), smoother finger-like protrusions form, resembling the invasive fingering  driven by apoptosis, which has been observed in viscous  models \cite{cristini2003,Pham2018}. 

These results highlight the interplay between mechanical resistance and apoptosis in determining tumor morphology. Shear elasticity favors buckling and sharp boundary deformations, whereas apoptosis promotes more invasive, smoother finger-like patterns. When both shear elasticity and apoptosis are present, our simulations consistently exhibit a superposition of these two mechanisms (Fig.~\ref{fig:instabilities} bottom right). The latter case is shown in Movie S2 (\Supp).
Notably, when apoptosis is present ($\lambda_a=0.1$), the addition of shear elasticity ($G_{\rm in}>0$) leads to strongly increased tumor size as compared to the pure invasive fingering case ($\lambda_a=0.1, G_{\rm in}=0$), highlighting how both instabilities cooperate to significantly accelerate the expansion of the total tumor area. 

\begin{figure*}[hbt!] 
    \centering
    \includegraphics[width=\textwidth]{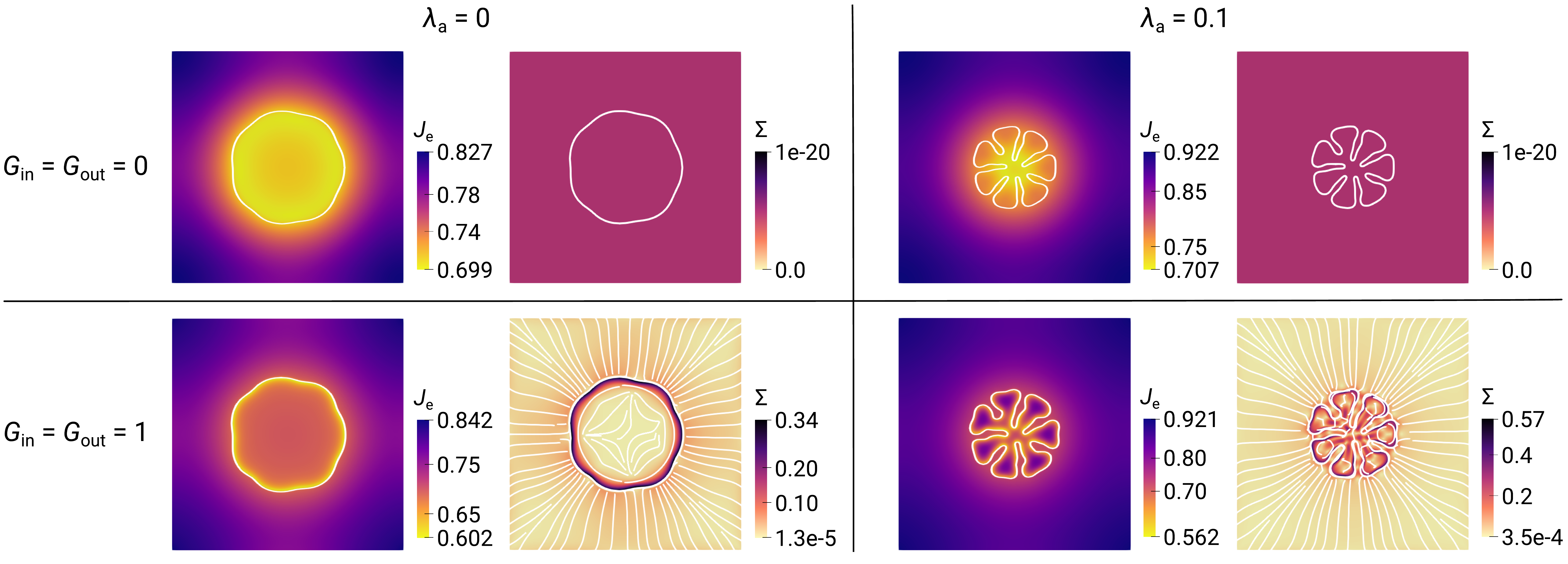}
    \caption{Tumor growth patterns at $t=240$ under varying mechanical and apoptotic conditions in a confined scenario ($\kappa\sub{out}>0, G\sub{out}\ge0$). Top row: $G\sub{in}=G\sub{out}=0$, bottom row: $G\sub{in}=G\sub{out}=1$; left column: $\lambda\sub{a}=0, \lambda\sub{p}=0.06$, right column: $\lambda\sub{a}=0.1, \lambda\sub{p}=0.2$. In the $\Sigma$ plots, streamlines indicate the direction of the eigenvector of the stress tensor corresponding to the smallest eigenvalue. Further simulation parameters: $L_x=L_y=16, h=0.04, R_0=0.8, \epsilon=0.045, \zeta=6.25, M=0.001, L= 0.3, \beta\sub{in} = \beta\sub{out}=0.3, \kappa\sub{in} = \kappa\sub{out} = 2$.}
    \label{fig:instabilities_constrained}
\end{figure*}

Interestingly, we observe a markedly different behavior when bulk elasticity is present in the surrounding tissue, see \figref{fig:instabilities_constrained}. 
While we still observe finger-like protrusions for $\lambda_a >0$, the elastic buckling for $G>0$ seems largely suppressed. 
This observation is consistent with the mechanical constraints of the system; buckling necessitates significant shear deformation of the surrounding medium, which is energetically penalized when the host tissue possesses a non-zero shear modulus.
Moreover, in the absence of shear elasticity and apoptosis, the tumor no longer grows isotropically; instead, slight protrusions develop at the surface. 
This new instability seems to be driven by the compression of outer tissue with a non-zero bulk modulus, as we demonstrate in detail in the \Supp ~(Sec. on the role of surrounding tissue in tumor surface instabilities; Fig. S3).

\subsection*{Parameter studies of unconfined growth}

In the subsequent sections, we analyze how the model parameters influence growth and protrusion formation in various unconfined and confined scenarios.
 
We first explore how variations in relaxation rate and compressibility affect the growth dynamics and morphology of an unconfined growing tumor ($\kappa\sub{out} = G\sub{out} = 0$). Fig.~\ref{fig:relaxation_study_unconstrained} (a) shows the evolution of shapes for three different values of the relaxation rate $\beta$. 
We find that smaller values of $\beta$ lead to enhanced growth of tumor protrusions. 
Obviously, a smaller relaxation rate constricts the material, leading to significantly stronger local material compression and dilatation (i.e. larger variation in $J_e$). This trend is also observed when comparing tumors of equal size for different $\beta$, see Fig.~\ref{fig:relaxation_study_unconstrained} (a). Conversely, an increased relaxation rate reduces the tumor surface area due to tissue rearrangement, thereby leading to slower growth.

Moreover, the compressibility of the growing material has an influence on protrusion formation. In Fig.~\ref{fig:relaxation_study_unconstrained} (b) we display simulation results of an almost incompressible material ($\kappa\sub{in}/G\sub{in}=100$). Here, incompressibility implies that, as a result of the high bulk modulus ($\kappa\gg G$), local material extension approaches the local growth rate, i.e. $\nabla\cdot{\bf v} = \gamma$.
Comparing Figs.~\ref{fig:relaxation_study_unconstrained} (a) and (b), we find that the incompressible material shows similar morphology but experiences less local compression as indicated by $J_e\approx 1$ in Fig.~\ref{fig:relaxation_study_unconstrained} (b). 
In both compressible and incompressible cases, larger tumor fluidity (i.e., large $\beta$) leads to slower growth as indicated by the different times when tumors reach the same area. 
We conclude that the growth of protrusions is increased by the presence of shear stresses (which are suppressed by $\beta$), consistent with the previous observation that shear stress drives elastic buckling (Fig.~\ref{fig:instabilities} bottom left). 

We now examine the impact of tissue compressibility $\kappa/G$ and fluidity $\beta$ on the overall growth rate of the tumor. 
At early times, compressible material grows faster than nearly incompressible material, Fig.~\ref{fig:relaxation_study_unconstrained} (e). 
The observed behavior is consistent with the results from the Eulerian nonlinear radially symmetric model \cite{wei2023eulerian} and was already explained as follows: 
As the nutrient concentration increases towards the tumor boundary, more cells proliferate there, resulting in higher cell density at the boundary and lower density at the center, see Figs.~\ref{fig:relaxation_study_unconstrained}(a, b) at early times. In 
compressible tissue (i.e. small $\kappa$), the cells expand in size as they move towards the center, leading to increase in overall tumor size. In contrast, in incompressible tissue, cells change their shape but not their size during tissue flow. That implies that weakly fluidized compressible tumors grow faster at early times.

Interestingly, for later times, we find the opposite relation, which was not yet reported in the radially symmetric simulations \cite{wei2023eulerian}: the incompressible tumor grows faster, even catching up the size of the compressible one, Fig.~\ref{fig:relaxation_study_unconstrained} (e). 
This can be explained by the fact that, at later times, tumor growth is driven exclusively by the outgrowth of protrusions. Correspondingly, there is much more tumor surface area as compared to the radially symmetric case of same tumor volume. Cells in this surface region are compressed due to local proliferation, reducing the size of the tumor in the compressible case. Opposed to that, in the incompressible tumor, cells immediately assume their desired volume as they grow, leading to a faster growth of the proliferating surface region. 
This is also evident from the higher velocity magnitude within the tumor, as depicted in Fig.~\ref{fig:relaxation_study_unconstrained} (d) compared to (c). 
In summary, compressibility of the tumor \textit{interior} leads to increased tumor size, while compressibility of the tumor \textit{boundary} leads to a decrease in tumor size. 

Moreover, dynamically reducing the tumor's bulk modulus, $\kappa\sub{in}$, i.e., switching from incompressible to compressible material, instantaneously decreases the rate of area change. Fig.~\ref{fig:relaxation_study_unconstrained} (g) shows that the rate of area change (dashed red curve) approaches that of the compressible material (purple curve).

\subsection*{Topology change}

\begin{figure*}[p]

\centering
\includegraphics[width=0.96\textwidth]{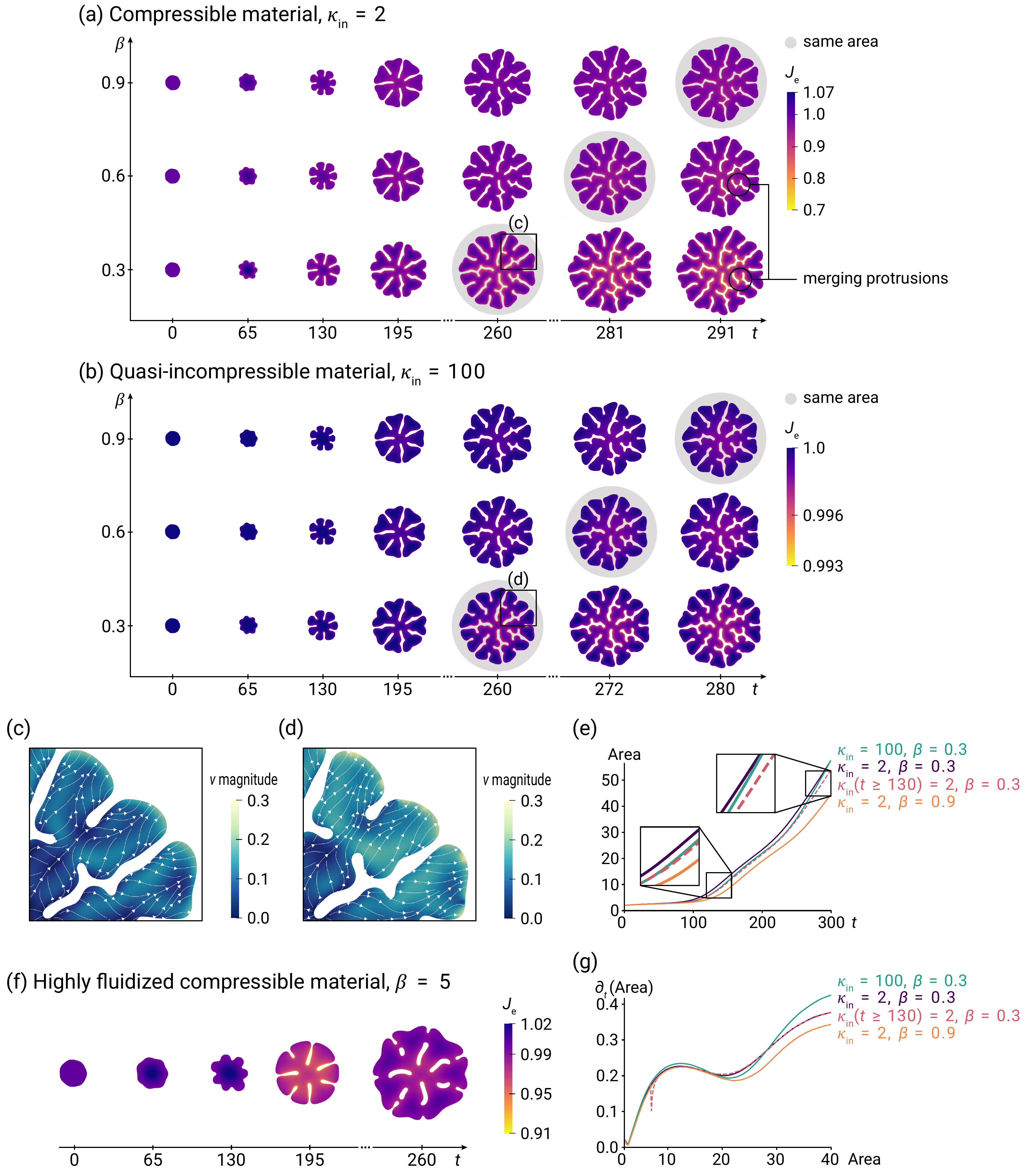}

\caption{Influence of fluidity on unconfined tumor evolution ($\kappa\sub{out}=G\sub{out}=0$) with perturbed initial condition (modes 2 and 7). Comparison of the shape evolution of (a) a compressible tumor, $\kappa\sub{in}=2, G\sub{in}=1$, and (b) a quasi-incompressible tumor, $\kappa\sub{in}=100, G\sub{in}=1$, depending on relaxation $\beta = \beta\sub{in} = \beta\sub{out}$. The gray-shaded shapes have approximately the same area ($\text{error } < 0.3 \%$). Below: Velocity fields of a section of (c) compressible and (d) quasi-incompressible tumor for $\beta = 0.3$ at $t=260$. The respective sections are marked in the above plots. (e) Tumor area as a function of time for different cases of compressibility and fluidity including dynamic change of the tumor's bulk modulus from $\kappa\sub{in}=100$ to $\kappa\sub{in}=2$ at $t=130$ (red curve). (f) Highly fluidized compressible tissue ($\beta=5, \kappa\sub{in}=2, G\sub{in}=1$) shows early changes in topology. (Only in this case we used a smaller computational domain with $L_x=L_y=6.4$.) (g) Rate of area change as a function of the area. Further simulation parameters: $L_x=L_y=12, h=0.04, R_0=0.8, \epsilon=0.045$, $\zeta=6.25$, $M=0.001$, $\lambda\sub{p} = 0.2$, $\lambda\sub{a} = 0.1$, $L= 0.3$.}
\label{fig:relaxation_study_unconstrained}
\end{figure*}

As noted above, topological changes of the tumor morphology are generally inhibited by shear stress of trapped material between the branching protrusions of the tumor, see also \Supp ~Fig. S2. 
However, we observe the merging of the tumor protrusions in the last two rows of Fig.~\ref{fig:relaxation_study_unconstrained} (a) between times $t=281$ and $t=291$ (see circled regions). 
Topological transitions are observed exclusively in compressible tumors, a phenomenon mechanistically attributable to the ability of compressible media to redistribute internal pressure while generating significantly lower shear stresses. 
Given that shear stress acts as the primary inhibitor of topological rearrangements (as detailed in the \Supp ~Sec. regarding the influence of the diffuse interface parameter), compressibility inherently facilitates the merging process. 
Similarly, our results indicate that high tissue fluidity promotes topological transitions by accelerating the relaxation of shear stresses that emerge during the coalescence of protrusions. 
Using an increased fluidity parameter ($\beta = 5$), as shown in Fig.~\ref{fig:relaxation_study_unconstrained} (f) and Movie S3, we observe multiple early-stage topological changes. 
Furthermore, we demonstrate in the \Supp (see Sec. on tissue fluidization at the tumor boundary) that localized increases in shear relaxation at the interface further facilitate these transitions.

\subsection*{Parameter studies of confined growth}
The phase-field model allows us to incorporate the mechanics of a surrounding tissue by taking non-zero values for $\kappa\sub{out}, G\sub{out}$ and $\beta\sub{out}$. In the following, we examine the effects of these parameters on the dynamics of growth and morphology.

\begin{figure*}[hbt!]
\centering
\includegraphics[width=\textwidth]{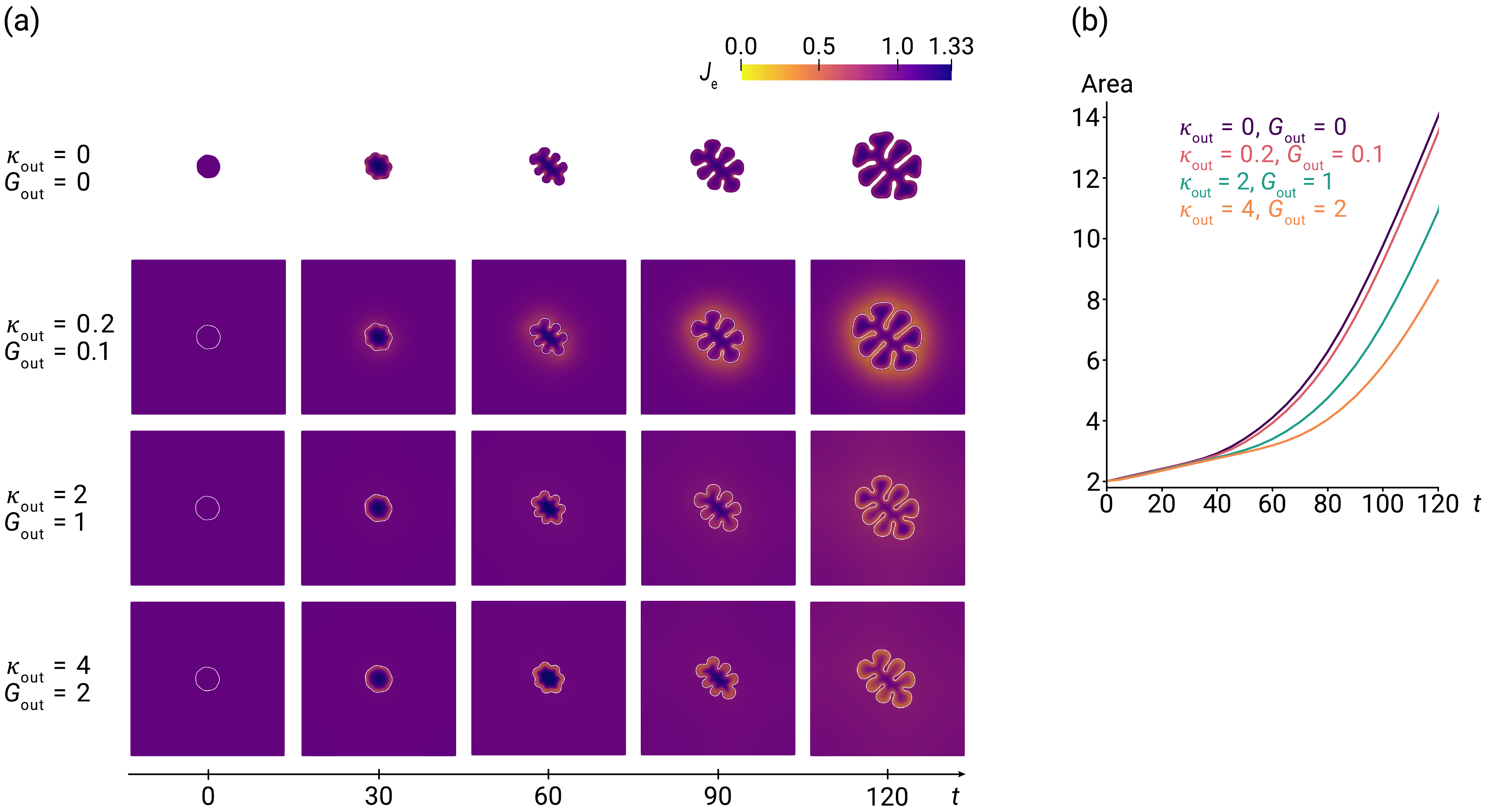}

\caption{(a) Protrusion formation for four cases of outer medium elastic parameters. In all cases, we assume compressible tumor material, $\kappa\sub{in} = 2, G\sub{in} =1$, with relaxation $\beta\sub{in} = 0.05, \beta\sub{out} = 0.2$. (b) Tumor area over time in the above four cases. (c) Rate of area change over area. Further simulation parameters: $L_x=L_y=10.4, h=0.04, R_0=0.8, \epsilon=0.045$, $\zeta=6.25$, $M=0.001$, $\lambda\sub{p} = 0.2$, $\lambda\sub{a} = 0.1$, $L= 0.3$.}
\label{fig:constrained_growth_perturbed}
\end{figure*}

Fig.~\ref{fig:constrained_growth_perturbed} presents a comparison of confined growth under different properties of the surrounding tissue. The first row shows the unconfined scenario with $\kappa\sub{out} = G\sub{out}=0$. With only a small elastic response ($\kappa\sub{out} = 0.2, G\sub{out}=0.1$), the outer medium exhibits strong local compression ($J_e\ll 1$) near the tumor, with the peak compression occurring between the outgrowing protrusions (second row). The increase of elastic moduli balances the elastic deformation in the outer medium (third row). To achieve this, the outer medium must resist compression in the regions around the protrusions, which constricts protrusions and reduces their length. 
A similar trend is observed when doubling $\kappa\sub{out}, G\sub{out}$ (fourth row), leading to slower outgrowth and shorter protrusions combined with greater compression of the tumor material close to the boundary. 
This behavior is also reflected in the evolution of the tumor area, as shown in Fig.~\ref{fig:constrained_growth_perturbed}(b). 
While the presence of the outer medium has little impact on growth in the early stages, it increasingly inhibits growth at later times due to the accumulation of compressive stress.

The confined growth model qualitatively reproduces the behavior observed in experiments. Studies involving different types of tumor spheroids growing in a 3D surrounding gel demonstrate sensitivity to the stiffness of their environment \cite{Helmlinger1997,Taubenberger2019, mahajan}. These studies showed that spheroids are significantly smaller and more compact (with a higher density of nuclei) when growing in a stiff microenvironment compared to a compliant one. Furthermore, it has been shown that in stiff hydrogels, a higher level of compressive stress builds up around the spheroids, and that cells stiffen when grown in stiffer hydrogels \cite{Taubenberger2019}.

\subsection*{3D result}
\begin{figure*}[hbt!]
\centering
\includegraphics[width=0.85\textwidth]{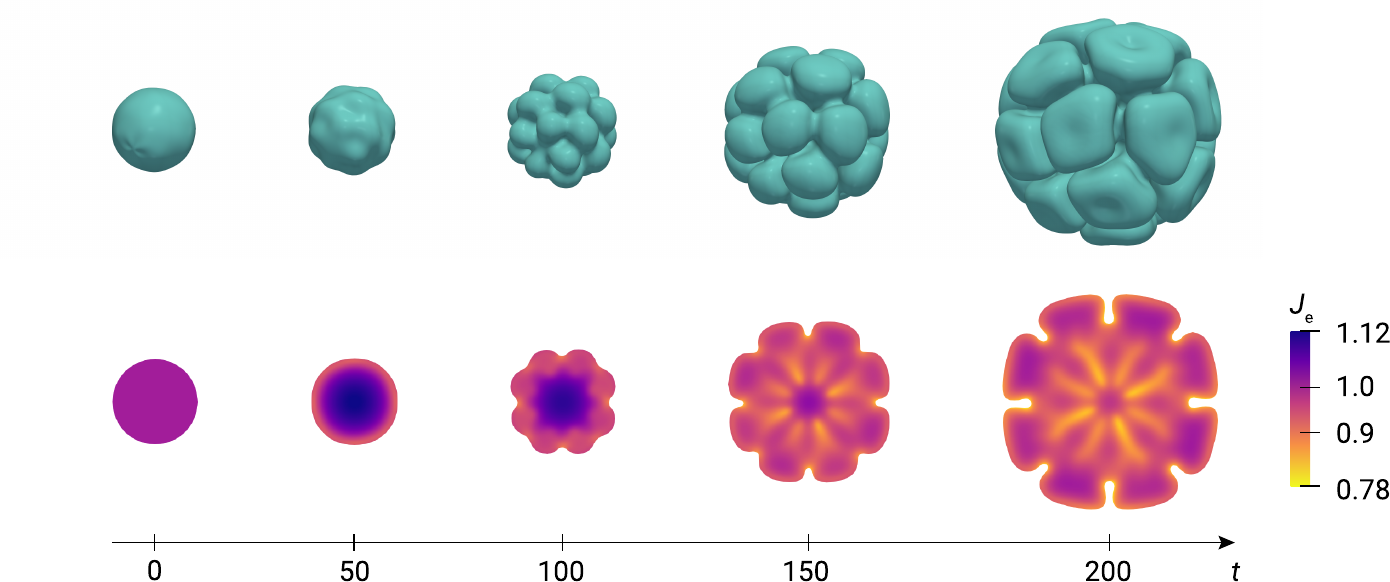}

\caption{Simulation of a 3D tumor exhibiting buckling behavior. Top row: Tumor contour in 3D, initially perturbed with a sinusoidal perturbation with frequency parameter $8$ along both polar angles. Bottom row: $J_e$ in the x-y plane through the tumor center. The material at the tumor boundary experiences increased compression, leading to enhanced growth at the surface. Simulation parameters: $L_x=L_y=L_z=7, h=0.05, R_0=1.3, \epsilon=0.05, \zeta=6.25, M=0.001, \lambda\sub{p} = 0.2, \lambda\sub{a} = 0.1, L= 0.3, \beta\sub{in} = \beta\sub{out}=0.3, \kappa\sub{in} = 2, G\sub{in} = 1, \kappa\sub{out} = G\sub{out} = 0.$ }
\label{fig:3dResult}
\end{figure*}

Finally, we illustrate the capability of our numerical framework to simulate differential growth in full 3D. The method described in the \Supp ~can be readily extended to three dimensions. The tumor is initialized as a spherical shape, with a sine perturbation added to its surface by calculating spherical coordinates $(\theta, \varphi)$ and applying a sinusoidal perturbation in $\theta$ and $\varphi$ to the tumor radius $R_0$. We observe a consistent outgrowth of the imposed undulations, akin to the 2D case, see Fig.~\ref{fig:3dResult} and Movie S4 (\Supp).

\section*{Conclusion}
We have developed a novel, mass-conservative model for the growth of compressible, hyperelastic tumors (and growing tissues more generally). 
A key component of our framework is the application of a phase-field approach to model the complex, mechanical interaction between two distinct material phases (tumor and surrounding host tissue). 
Growth within the tumor is prescribed by a diffusible growth factor, establishing a feedback loop where shape changes directly modify the local growth rate. We validated the phase field model by comparing the results with those from a sharp interface model \cite{Olaranont2024} and demonstrating convergence as the diffuse interface width $\epsilon\rightarrow 0$.

In the 1D radially symmetric setting, differential growth with a diffusible factor leads to stable, stationary tumor shapes, consistent with established literature. 
However, in 2D and 3D, these stationary solutions are  unstable, leading to symmetry breaking and complex morphogenesis. 
We identified two primary mechanisms responsible for this shape evolution, whose non-linear interplay leads to strongly branched tumor morphology: (i) an elastic buckling instability driven by compressive tangential stress along the tumor boundary, and (ii) invasive fingering resulting from differential growth coupled with apoptosis. 
Notably, while the elastic buckling phenomenon is driven by shear stresses \textit{within} the tumor, it is suppressed by shear stresses \textit{outside} the tumor when a host tissue of non-zero shear modulus is included.
Furthermore, we demonstrated that when combined, these two instabilities cooperate to significantly accelerate the expansion of the total tumor area. 

The morphogenesis of these complex structures can occur without topological transitions, as high internal shear stresses inhibit the close contact between different parts of the tumor boundary. However, if these shear stresses are sufficiently relaxed -- either by increasing tissue fluidity ($\beta$) or by the compressibility of the tumor material -- topological transitions, such as the merging of protrusions, are readily observed.

Concerning overall growth dynamics, our model shows that compressible tumors initially grow faster, consistent with previous findings for radially symmetric tumors \cite{wei2023eulerian}. However, this trend is inverted following symmetry breaking: the resulting increase in the tumor surface-to-volume ratio, combined with the mechanical effects of tumor compressibility, leads to a decrease in the overall volumetric growth rate at later times.

In summary, our study provides a robust theoretical framework demonstrating how the interplay between tissue mechanics and growth factor signaling dictates the intricate morphological stability and invasive potential of solid tumors.
Future work could focus on incorporating dynamic changes in the host tissue, such as matrix remodeling or degradation, or exploring the effects of stromal and  immune cell interactions on the tumor biomechanical microenvironment to further bridge the gap between theoretical modeling and clinical observations. 

\section*{Author Contributions}
L.Z., M.W., J.L. and S.A. developed the mathematical model and designed the study;
L.Z. discretized and implemented the system, performed all simulation studies, processed, analyzed, and visualized the data;
L.Z., M.W., J.L. and S.A. discussed the results and wrote the manuscript;
C.W. provided benchmark data;
S.A. funded and supervised L.Z. 
All authors approved the final version.

\section*{Acknowledgements}
SA acknowledges support from the German Research Foundation (grant 511509575). JL acknowledges partial support from the US National Institutes of Health through grants P30 CA062203 for the Chao Family Comprehensive Cancer Center at UC Irvine and P01CA288663 as well as the US National Science Foundation (grant DMS-1763272) and the Simons Foundation (grant 594598QN) for an NSF-Simons Center for Multiscale Cell Fate Research. MW acknowledges partial support from the US National Science Foundation (grants DMS-2012330, DMS-2144372). CW acknowledges partial funding from National Natural Science Foundation of China under grants 12371392 and 12431015. LZ and SA gratefully acknowledge computing time on the high-performance computer at the NHR Center of TU Dresden. This center is jointly supported by the Federal Ministry of Education and Research and the Free State of Saxony governments participating in the NHR (www.nhr-verein.de/unsere-partner).
This research was supported in part by grant NSF PHY-2309135 and the Gordon and Betty Moore Foundation Grant No. 2919.02 to the Kavli Institute for Theoretical Physics (KITP).

\bibliographystyle{abbrvnat}
\bibliography{references}

\balancecolsandclearpage

\onecolumngrid


\clearpage
\begin{center}
{\Large \textbf{Supplementary Information}}\\[1em]
\end{center}

\setcounter{figure}{0}
\renewcommand{\thefigure}{S\arabic{figure}}


\section*{Numerical method (2D, 3D)}
\label{sec:numerical_method_2d}

In this section, we present the numerical method used to solve the model in two and three dimensions. Maintaining a stable, linear, monolithic coupling of the system becomes increasingly challenging in higher dimensions. In full 3D, the linearization of strain evolution alone results in hundreds of terms, making the implementation complex and prone to errors.

To address these challenges, we employ an explicit Finite Difference Method. While this approach requires small time step sizes, it enables efficient parallelization.

\subsection*{Time discretization}
At the beginning of each time step, the stress is computed from the values of the left Cauchy-Green tensor from the preceding iteration,
\begin{align*}
    \mathbf{S}^{n-1} &= \kappa(\phi^{n-1}) ({J_e}^{n-1} - 1)\mathbf{I} + G(\phi^{n-1}) ({J_e}^{n-1})^{-\frac{2+d}{d}}\left(\mathbf{B_e}^{n-1}-\frac{1}{d}\text{tr}(\mathbf{B_e}^{n-1})\mathbf{I}\right) 
\end{align*}
with
\begin{equation*}
    {J_e}^{n-1} = \sqrt{\det (\mathbf{B_e}^{n-1})}. 
\end{equation*}
This is followed by updating the velocity,
\begin{align*}
    \zeta \mathbf{v}^{n-1} &= \nabla \cdot \mathbf{S}^{n-1}.
\end{align*}
At every time step we compute the solution of the concentration equation 
\begin{align}
\begin{array}{r l}
     - L^2 \Delta c^n + c^n = 0 &  \text{ in } \Omega_\text{in}^n,\\
     c^n = 1 &  \text{ on } \Gamma^n.
\end{array} 
\end{align}
Afterwards, the explicit Euler scheme for the evolution equation for $\mathbf{B_e}$ is
\begin{align}
    \frac{\mathbf{B_e}^{n}-\mathbf{B_e}^{n-1}}{\Delta t} &+ \mathbf{v}^{n-1}\cdot \nabla \mathbf{B_e}^{n-1} - \left(\nabla \mathbf{v}^{n-1}\right) \cdot \mathbf{B_e}^{n-1} - \mathbf{B_e}^{n-1} \cdot \left(\nabla \mathbf{v}^{n-1}\right)^\top \nonumber \\
    &= - \frac{2}{d} \gamma(\phi^{{n-1}},c^{n}) \mathbf{B_e}^{n-1} - 2 \beta\left(\phi^{n-1}\right) \mathbf{B_e}^{n-1} \left( \mathbf{B_e}^{n-1}  - \frac{\text{tr}\left(\mathbf{B_e}^{n-1}\right)}{d} \mathbf{I}\right). \label{eq:strain_time_discrete}
\end{align}
Finally, the  Cahn-Hilliard equation is solved by an explicit Euler scheme,
\begin{align}
    \frac{\phi^{n}-\phi^{n-1}}{\Delta t} + \mathbf{v}^{n-1} \cdot \nabla \phi^{n-1} &= M \Delta \mu^{n-1}, \label{eq:ch_time_discrete_phi}\\
    \mu^{n-1} &= f'(\phi^{n-1}) - \epsilon^2 \Delta \phi^{n-1}. \label{eq:ch_time_discrete_mu}
\end{align}
The presence of a fourth-order spatial derivative introduces a time step restriction of the form $\Delta t < C h^4/(M\epsilon^2)$, where $h$ is the width of the spatial grid. Fortunately, this limitation becomes less restrictive when considering that aiming to approach the sharp interface limit, leads to choosing the interface thickness $\epsilon$ on the order of the grid size. Consequently, the time step restriction can effectively be reduced to $\Delta t < Ch^2/M$, which is not too severe for small mobilities.

\subsection*{Space discretization}

We discretize the previously presented time discrete scheme in space using a Finite Difference Method. We subdivide the domain by an equidistant rectangular grid with equal spacing $h$ in all directions. For simplicity, we present the procedure for the two-dimensional problem. The scheme can easily be extended to three dimensions. In the main text, we will present simulation results for the two-dimensional as well as for the three-dimensional case. Let the computational domain $\Omega = [0, L_x] \times [0, L_y]$ be subdivided into $N_x \cdot N_y$ square cells with $L_x = N_x \cdot h$ and $L_y = N_y \cdot h$, respectively. 

We employ a staggered Cartesian grid in order to avoid spatial oscillations.
Thus, the velocity variables $v_i, i=1,2$, are shifted by $h/2$ in their respective direction onto the cell faces. All other variables are defined at the centers of the cells. Around the boundary facets, an additional layer of ghost points is added. For the sake of readability, we use integer indices to reference the values of the variables. Hence, the discrete values of the velocity field are
\begin{align*}
    {v_0}_{i,j} &= v_0\left((i-1)h, (j- 1/2)h\right), & i = 0, ..., N_x+2, j = 0,..., N_y+1, \\
    {v_1}_{i,j} &= v_1\left((i-1/2)h, (j- 1)h\right), & i = 0, ..., N_x+1, j = 0,..., N_y+2, 
\end{align*}
whereas the discrete values of central variables $q(x,y)$ are defined as
\begin{align*}
    {q}_{i,j} &= q\left((i- 1/2)h, (j- 1/2)h\right), & i = 0, ..., N_x+1, j = 0,..., N_y+1. \\
\end{align*}

We introduce the following discrete operators for some quantity $q$: the discrete Laplacian $\Delta_h$, horizontal and vertical averages $A_x, A_y$, backward differences $D_{x}, D_{y}$, and central differences $D_{2x}, D_{2y}$.
\begin{align*}
    \Delta_h q_{i,j} &:= (q_{i-1,j}+ q_{i+1,j} + q_{i,j-1} + q_{i,j+1} - 4q_{i,j})/h^2, \\
    A_x q_{i,j} &:= (q_{i,j}+q_{i-1,j})/2,\\
    A_y q_{i,j} &:= (q_{i,j}+q_{i,j-1})/2, \\
    D_{x} q_{i,j} &:= (q_{i,j}-q_{i-1,j})/h,\\
    D_{y} q_{i,j} &:= (q_{i,j}-q_{i,j-1})/h, \\
    D_{2x} q_{i,j} &:= (q_{i+1,j}-q_{i-1,j})/(2h),\\
    D_{2y} q_{i,j} &:= (q_{i,j+1}-q_{i,j-1})/(2h).
\end{align*}
When the discrete operators are applied to tensor-valued variables, they are evaluated component-wise. In order to compute derivatives of a central variable at a cell face center, we need a modified version of the central difference operator. The arguments of the central difference are averages along the orthogonal axis,
\begin{align*}
    \tilde{D}_{2x} q_{i,j} &:= D_{2x} A_y q_{i,j}, \\
    \tilde{D}_{2y} q_{i,j} &:= D_{2y} A_x q_{i,j}.
\end{align*}

The space discretization of the stress reads
\begin{align}
    \mathbf{S}_{i,j}^{n-1} &= \kappa\left(\phi_{i,j}^{n-1}\right) \left({J_e}_{i,j}^{n-1} - 1\right)\mathbf{I} + G\left(\phi_{i,j}^{n-1}\right) \left({J_e}_{i,j}^{n-1}\right)^{-\frac{5}{3}}\left(\mathbf{B_e}_{i,j}^{n-1}-\frac{1}{3}\text{tr}\left(\mathbf{B_e}_{i,j}^{n-1}\right)\mathbf{I}\right) \label{eq:stress_space_discrete} \\
    & \text{ for } i=1,..., N_x+1, j=1,..., N_y+1, \nonumber
\end{align}
with
\begin{equation*}
    {J_e}_{i,j}^{n-1} = \sqrt{\det \left(\mathbf{B_e}_{i,j}^{n-1}\right)}. 
\end{equation*}

Having updated the stress $\mathbf{S}_{i,j}^{n-1}$, the velocity $\mathbf{v}_{i,j}^{n-1}$ can be computed from 
\begin{align}
    {v_0}_{i,j}^{n-1} &= \frac{1}{\zeta} \left( D_x\left({\mathbf{S}_{xx}}_{i,j}^{n-1}\right) + \tilde{D}_y \left({\mathbf{S}_{yx}}_{i,j}^{n-1}\right)\right) \\
    & \text{ for } i=1,..., N_x+1, j=1,..., N_y, \nonumber \\
    {v_1}_{i,j}^{n-1} &= \frac{1}{\zeta} \left( \tilde{D}_x\left({\mathbf{S}_{xy}}_{i,j}^{n-1}\right) + D_y \left({\mathbf{S}_{yy}}_{i,j}^{n-1}\right)\right) \\
    & \text{ for } i=1,..., N_x, j=1,..., N_y+1. \nonumber
\end{align}

In the next step the nutrient concentration field is calculated. We discretize the nutrient equation in the tumor interior as
\begin{align}
   L^2 \Delta_h c_{i,j}^{n} - c_{i,j}^{n} &= 0 
    & \text{ for } i=1,..., N_x, j=1,..., N_y \text{ with }\phi_{i,j}^n>0. \label{discrete nutrient}
\end{align}
Since the concentration changes only slightly in every time step, we perform only a single Jacobi step of this linear system, reducing computational effort to $N_x\cdot N_y$ operations. 
To guarantee the boundary condition $c=1$ at the tumor boundary, we use an embedded boundary approach \cite{johansen1998,balajewicz2014reduction}, which will be explained in the following section. 

After that, we compute $\mathbf{B_e}$ from the discrete form of \eqref{eq:strain_time_discrete},
\begin{align}
    \mathbf{B_e}_{i,j}^{n}&= \mathbf{B_e}_{i,j}^{n-1} +
    \Delta t \left( 
    - A_x {v_0}_{i+1,j}^{n-1}\cdot D_{2x} \mathbf{B_e}_{i,j}^{n-1} 
    - A_y {v_1}_{i,j+1}^{n-1}\cdot D_{2y} \mathbf{B_e}_{i,j}^{n-1}  \right) \nonumber \\
    &+\Delta t \left( 
    \left(\nabla_h \mathbf{v}_{i,j}^{n-1}\right) \cdot \mathbf{B_e}_{i,j}^{n-1} 
    + \mathbf{B_e}_{i,j}^{n-1} \cdot \left(\nabla_h \mathbf{v}_{i,j}^{n-1}\right)^\top \right) \nonumber \\
    &- \frac{2}{3} \Delta t \gamma\left(\phi_{i,j}^{n-1},c_{i,j}^{n}\right)\mathbf{B_e}_{i,j}^{n-1} - 2 \Delta t \beta\left(\phi_{i,j}^{n-1}\right) \mathbf{B_e}_{i,j}^{n-1} \left( \mathbf{B_e}_{i,j}^{n-1}  - \frac{1}{3} \text{tr}\left(\mathbf{B_e}_{i,j}^{n-1}\right)\mathbf{I}\right)  \\
    & \text{ for } i=1,..., N_x, j=1,..., N_y, \nonumber
\end{align}
where the discrete Jacobian is given by 
\begin{equation*}
    \nabla_h \mathbf{v}_{i,j}^{n-1} := 
    \begin{pmatrix} 
    D_x {v_0}_{i+1,j}^{n-1} & \tilde{D}_{2y} {v_0}_{i+1,j}^{n-1} \\
    \tilde{D}_{2x} {v_1}_{i,j+1}^{n-1} & D_y {v_1}_{i,j+1}^{n-1}
    \end{pmatrix}.
\end{equation*}

After updating $\mathbf{B_e}$, the Cahn-Hilliard system is solved. Employing central differences in the advective term leads to a second order space discretization of the Cahn-Hilliard equation \eqref{eq:ch_time_discrete_phi} and \eqref{eq:ch_time_discrete_mu} of the form
\begin{align}
    \mu_{i,j}^{n-1} &= f'\left(\phi_{i,j}^{n-1}\right) - \epsilon^2 \Delta_h \phi_{i,j}^{n-1}, \\
    \phi_{i,j}^{n} &= \phi_{i,j}^{n-1}  - \Delta t \left( A_x {v_0}_{i+1,j}^{n-1} D_{2x} \phi_{i,j}^{n-1} + A_y {v_1}_{i,j+1}^{n-1} D_{2y} \phi_{i,j}^{n-1} \right) + M \Delta t \Delta_h \mu_{i,j}^{n-1} \label{eq:phi_space_discrete} \\
    & \text{ for } i=1,..., N_x, j=1,..., N_y. \nonumber
\end{align}

For finite $\epsilon$, the phase field may deviate from the values of $1$ in the tumor and $-1$ in the outer medium during its evolution. To account for this deviation, the phase-dependent parameters $\kappa$, $G$, and $\beta$ are defined using the truncated variant of $\phi$,
\begin{equation*}
    \tilde\phi_{i,j}^{n-1} = \max\left\{\min\left\{\frac{2}{1.8}\phi_{i,j}^{n-1}, 1 \right\}, -1 \right\}.
\end{equation*}
This formulation maps $\phi$-values of $\pm 0.9$ to $\tilde\phi=\pm 1$, ensuring that the correct parameters are used within each phase, even if the phase field does not assume the extremal values of $\pm 1$ in the phases. This is in particular necessary to deal with large parameter contrasts between the two phases, for example if one phase is inelastic ($\kappa=G=0$) while the other is elastic. 
Finally, homogeneous Neumann boundary conditions apply to the phase field and to $\mathbf{B_e}$ at the domain boundary. They are incorporated at the end of each time step. 

Equations \eqref{eq:stress_space_discrete} - \eqref{eq:phi_space_discrete} offer a direct way for solving the system. In particular, these equations can be solved successively and independently, eliminating the need to solve a coupled system. As a result, the solution algorithm can be implemented and parallelized straightforwardly. We leverage the OpenMPI application programming interface to execute the code on shared-memory systems.

\subsection*{Embedded boundary approach} \label{sec:ebm}

An accurate discretization of the nutrient concentration equation requires the boundary condition $c=1$ to hold on the tumor boundary $\Gamma$. Given by the $0$-level set of $\phi$, the interface $\Gamma$ is not fitted to the spatial grid which impedes a direct implementation of this condition.
Instead, we adopt an embedded boundary approach \cite{johansen1998,balajewicz2014reduction}. The
concentration values in a narrow band of ghost points outside the tumor are chosen such that a linear interpolation between the ghost point and its neighboring interior points consistently yields $c=1$ on $\Gamma$. 

The precise algorithm is illustrated below in 2D. The extension to 3D is straightforward. 
The signed distance to $\Gamma$ is first computed based on the phase field, $\mathrm{dist}_{i,j}=-\sqrt{2}\epsilon~\text{atanh}(\phi(x_i,y_j))$ in every grid point. 
This function is negative inside $\Gamma$ and positive outside $\Gamma$. 
The discrete nutrient equation \eqref{discrete nutrient} is only solved on grid points inside of $\Gamma$, i.e. where $\mathrm{dist}_{i,j}<0$.
Since the discrete Laplacian operator couples interior points near $\Gamma$ with their exterior neighbors, we must compute the values $c_{i,j}^n$ at these ghost points prior to solving Eq.~\eqref{discrete nutrient}.
To this end, the subsequent procedure is performed on all ghost points $(i,j)$ with $0 < \mathrm{dist}_{i,j} < h$.
First, compute the normal vector 
\begin{align*}
({\bf n})_{i,j} = \frac{(D_{2x}\mathrm{dist}_{i,j}, D_{2y}\mathrm{dist}_{i,j})}{||(D_{2x}\mathrm{dist}_{i,j}, D_{2y}\mathrm{dist}_{i,j})||_2}.
\end{align*}
The concentration value $\tilde{c}^n_{i,j}$ for the ghost point $(i,j)$ is then computed via a linear extrapolation from its neighboring interior grid points $(i', j')$ at the previous time step $n-1$,
\begin{align} \label{eq: embedded ghost point update}
\tilde{c}^{n}_{i,j} = 1 + \sum_{(i',j')\in\mathcal{N}(i,j)} \frac{\mathrm{dist}_{i,j}}{\mathrm{dist}_{i',j'}}  \left({\bf n}_{i,j} \cdot
\left(
\begin{array}{c}
     i-i'\\
     j-j'
\end{array}\right)
\right)^2 (c^{n-1}_{i',j'}-1)
\end{align}
where $\mathcal{N}(i,j)$ are the interior Cartesian neighbors of $(i,j)$. 
The quadratic term represents a normal projection which weights the contributions of $x$ and $y$ neighbors according to the $\mathbf{n}$ vector.  
In the implementation, the magnitude of $\mathrm{dist}_{i,j} / \mathrm{dist}_{i',j'}$ is capped to $10$, to avoid ghost points of very large magnitude when an interior neighbor is very close to $\Gamma$.

Finally, to stabilize the fully coupled system, we apply an under-relaxation step, updating only an $\omega$-portion of $c$, with a small constant $\omega=0.1$, 
\begin{align} \label{eq: embedded relaxation}
c^{n}_{i,j} = (1 - \omega) c^{n-1}_{i,j}+ \omega \tilde{c}^n_{i,j}.
\end{align}
This update of ghost point values given in Eqs.~\eqref{eq: embedded ghost point update}-\eqref{eq: embedded relaxation} is performed once before solving (one Jacobi step of) the nutrient equation  \eqref{discrete nutrient}. 

\section*{1D radially symmetric model}
\label{sec:model_1D}

In this section, we introduce a radially symmetric representation of the system presented in the main text. We express the system of equations in spherical coordinates $(r,\theta, \varphi)$, where $r$ represents the radial distance, and $\theta$ and $\varphi$ are the angular coordinates. In the radially symmetric case, the solutions depend only on the radial coordinate $r$ and time $t$, reducing the complexity of the system while preserving the essential physical characteristics and dynamics of tissue growth.

This symmetry reduces the number of independent components of stress and strain to two: the radial and tangential components. The spatial derivatives also simplify; radial derivatives remain unchanged, while tangential derivatives are reduced due to the uniformity in the angular directions, resulting from the spherical symmetry.

Dirichlet boundary conditions ${\bf v}=0$ are imposed on the velocity at the tumor center $r=0$ and at the domain boundary, $r=R_\Omega$, where $R_\Omega$ is the fixed radius of the computational domain. 
The tumor boundary at any given time $t$ is defined by the radial position $r$ where the phase-field variable $\phi(r,t)=0$, representing the interface between the tumor and the surrounding tissue.
In the radially symmetric setting, the equation for the nutrient concentration $c$ is reformulated using a phase-field approach \cite{LiLowengrubRätzVoigt}.

With these assumptions and boundary conditions, the governing equations for spherical tumor growth in $d$ dimensions take the form 

\begin{align*}
    \partial_t \phi + v_r \cdot \partial_r \phi &= M \left( \frac{d-1}{r} \partial_r \mu + \partial_{rr} \mu \right), \\
    \mu &= f'(\phi) - \epsilon^2 \left( \frac{d-1}{r} \partial_r \phi + \partial_{rr} \phi \right), \\
    f(\phi) &= \frac{1}{4} (\phi^2 - 1)^2, \\
    \partial_t B_{rr} + v_r \partial_r B_{rr} - 2 B_{rr} \partial_r v_r &= -\frac{2}{d} \frac{\phi+1}{2} \gamma B_{rr} - 2 \beta(\phi) B_{rr} \left(B_{rr} - \frac{B_{rr} + (d-1)B_{\theta\theta}}{d} \right), \\
    \partial_t B_{\theta\theta} + v_r \partial_r B_{\theta\theta} - 2 B_{\theta\theta} \frac{1}{r} v_r &= -\frac{2}{d} \frac{\phi+1}{2} \gamma B_{\theta\theta} - 2 \beta(\phi) B_{\theta\theta} \left(B_{\theta\theta} - \frac{B_{rr} + (d-1)B_{\theta\theta}}{d} \right), \\
    \gamma &= \lambda\sub{p} c - \lambda\sub{a}, \\
    (1+\phi)c &= L^2 \left(\partial_r \phi \partial_r c + (1+\phi)\left(\frac{d-1}{r}\partial_r c + \partial_{rr} c  \right)\right)  +  (1-\phi) \frac{L^2}{\epsilon^2}(1-c), \\
    v_r &= \frac{1}{\zeta} \left(\partial_r S_{rr} + \frac{d-1}{r} (S_{rr} - S_{\theta\theta}) \right), \\
    S_{rr} &= \kappa(\phi) (J_e-1) + G(\phi) J_e^{-\frac{2+d}{d}}\left(B_{rr} - \frac{1}{d} (B_{rr} + (d-1)B_{\theta\theta})\right), \\
    S_{\theta\theta} &= \kappa(\phi) (J_e-1) + G(\phi) J_e^{-\frac{2+d}{d}}\left(B_{\theta\theta} - \frac{1}{d} (B_{rr} + (d-1)B_{\theta\theta})\right), \\
    J_e &= \sqrt{B_{rr}B_{\theta\theta}^{d-1}}.
\end{align*}


\section*{Numerical method (1D radially symmetric)}
\label{sec:numeric FEM}
After establishing the system of equations, it remains to solve the strongly coupled equations. This section explains the Finite Element Method that we use to solve the 1D radially symmetric problem in each time step. It is based on the Finite Element toolboxes DUNE \cite{sander2020} and AMDiS \cite{praetorius}. With the assumption of radial symmetry, computation reduces to only one dimension. 

First, we adopt a semi-implicit time discretization for solving the coupled system in the n-th time step. Let the time interval $\left[0, T\right]$ be divided in $N$ subintervals of size $\Delta t$, such that the n-th time step is $t^n := n \cdot \Delta t$. For the space discretization we use Lagrange-P2 elements for all variables, namely the phase-field variable $\phi$, the chemical potential $\mu$, the velocity $v_r$, the concentration $c$, the stresses $S_{rr}$ and $S_{\theta\theta}$ and the strains $B_{rr}$ and $B_{\theta\theta}$. 

We will present the complete discrete system of equations in the weak form. The finite element space
\begin{equation*}
    C_h = \left\{ c \in C^0(\bar{\Omega}) \mid c|_k \in P_2(k), k \in T_\Omega \right\}
\end{equation*}
is used, where $T_\Omega$ is the triangulation of the one dimensional domain $\Omega$.

To enhance readability, we choose to present the equations individually, addressing phase separation, strain, nutrient concentration, and elastic forces, even though they are part of a unified coupled system.

The Cahn-Hilliard equation is discretized implicitly in time, leading to the weak form: Find $(\phi^n, \mu^n) \in C_h \times C_h$ such that for all $(w_\phi, w_\mu) \in C_h \times C_h$ holds
\begin{align*}
    \int_{\Omega} \frac{\phi^n - \phi^{n-1}}{\Delta t}w_\phi \diff r &= - \int_{\Omega} \left( v^{n} \cdot \partial_r \phi^n \right) w_\phi \diff r
    - M \int_\Omega \partial_r \mu^n \partial_r w_\phi \diff r, \\
    \int_\Omega \mu^n w_\mu \diff r &= \int_\Omega \left((\phi^{n})^2-1\right)\phi^n w_\mu \diff r
    + \epsilon^2 \int_\Omega \partial_r \phi^n \partial_r w_\mu \diff r. 
\end{align*}

Note that, for the Cahn-Hilliard equation, no axisymmetric terms are necessary, as the phase field only tracks the interface and we do not simulate phase separation. The Cahn-Hilliard system is directly coupled to the elastic force system through the presence of $v_r^n$. The nonlinear terms in the Cahn-Hilliard equation are linearized using a first-order Taylor approximation, but we refrain from explicitly writing out the linearized terms for the sake of readability.

Numerical and computer arithmetic reasons may result in a minimal deviation of $\phi$ from its bounds. Therefore, we constrain $\phi$ to the interval $[-1,1]$ for subsequent calculations. Let 
\begin{equation*}
    \tilde{\phi} = \max\left\{ -1, \min\left\{ \phi, 1 \right\} \right\}.
\end{equation*}

We use the semi-implicit time discrete form of the strain equation taking $B_{rr}^{n-1}, B_{\theta\theta}^{n-1}, \tilde{\phi}^{n-1}$ and $c^{n-1}$  from the previous time step, but the velocity from the current time step. The weak form reads: Find $(B_{rr}^n, B_{\theta\theta}^n) \in C_h \times C_h$ such that for all $(w_1, w_2) \in C_h \times C_h$ holds
\begin{align*}
    \int_\Omega \frac{B_{rr}^n - B_{rr}^{n-1}}{\Delta t}w_1 \diff r &=  \int_\Omega \left(- v_r^n\partial_rB_{rr}^{n-1} +  2B_{rr}^{n-1}\partial_r v_r^n \right) w_1 \diff r + \int_\Omega-\frac{2}{d} \left(\lambda\sub{p} c^{n-1} - \lambda\sub{a}\right)\frac{\tilde{\phi}^{n-1}+1}{2} B_{rr}^{n-1} w_1 \diff r \\
    &\hspace{1em}- \int_\Omega  \frac{2(d-1)}{d} \beta\left(\tilde{\phi}^{n-1}\right) (B_{rr}^{n-1})^2  w_1 \diff r + \int_\Omega \frac{2(d-1)}{d}\beta\left(\tilde{\phi}^{n-1}\right) B_{rr}^{n-1}B_{\theta\theta}^{n-1} w_1 \diff r, \\
    \int_\Omega \frac{B_{\theta\theta}^n - B_{\theta \theta}^{n-1}}{\Delta t}w_2 \diff r &=  \int_\Omega \left(- v_r^n\partial_rB_{\theta\theta}^{n-1} + \frac{2}{r}B_{\theta \theta}^{n-1} v_r^n \right) w_2 \diff r + \int_\Omega-\frac{2}{d} \left(\lambda\sub{p} c^{n-1} - \lambda\sub{a}\right)\frac{\tilde{\phi}^{n-1}+1}{2} B_{\theta \theta}^{n-1} w_2 \diff r \\
    &\hspace{1em}- \int_\Omega \frac{2}{d} \beta\left(\tilde{\phi}^{n-1}\right) (B_{\theta\theta}^{n-1})^2  w_2 \diff r + \int_\Omega \frac{2}{d}\beta\left(\tilde{\phi}^{n-1}\right) B_{rr}^{n-1}B_{\theta\theta}^{n-1} w_2 \diff r, \\
    \text{with } \beta\left(\tilde{\phi}^{n-1}\right) &= \beta\sub{in}\frac{1+\tilde{\phi}^{n-1}}{2}  + \beta\sub{out} \frac{1-\tilde{\phi}^{n-1}}{2}.
\end{align*}

The semi-implicit time discretization of the nutrient concentration equation uses $\tilde{\phi}^{n-1}$ from the previous time step, but the concentration $c^n$ from the current time step. The weak form reads: Find $c^n \in C_h$ such that for all $w_c \in C_h$ holds
\begin{align*}
    \int_\Omega \left(1+\tilde{\phi}^{n-1}\right)c^n w_c \diff r = - &L^2 \int_\Omega \left(-\frac{d-1}{r}\partial_r c^n \left(1+\tilde{\phi}^{n-1}\right)  + \partial_r c^n \left(1+\tilde{\phi}^{n-1}\right) \partial_r \right) w_c \diff r \\
    + &L^2 \int_\Omega \left(1-\tilde{\phi}^{n-1}\right) \frac{1}{\epsilon^2}\left(1-c^n\right)w_c \diff r. 
\end{align*}

The velocity equation is discretized fully implicitly in time. The stress equations, on the other hand, are semi-implicitly discretized using $\tilde{\phi}$ from the previous time step alongside current stresses and strains. The weak form reads: Find $(v_r^n, S_{rr}^n, S_{\theta\theta}^n) \in C_h \times C_h \times C_h$ such that for all $(w_v, w_3, w_4) \in C_h \times C_h \times C_h$ holds
\begin{align*}
    \int_\Omega v_r^n w_v \diff r &= \int_\Omega \frac{1}{\zeta}\left(\partial_r S_{rr}^n + \frac{d-1}{r} S_{rr}^n - \frac{d-1}{r} S_{\theta\theta}^n\right)w_v \diff r, \\
    \int_\Omega S_{rr}^nw_3 \diff r &= \int_\Omega \kappa\left(\tilde{\phi}^{n-1}\right) \left(\left(B_{rr}^n\right)^{1/2} \left(B_{\theta\theta}^n\right)^{(d-1)/2} -1\right)w_3 \diff r \\
    &+ \int_\Omega G\left(\tilde{\phi}^{n-1}\right) \left(B_{rr}^{n}\left(B_{\theta\theta}^{n}\right)^{d-1}\right)^{-\frac{2+d}{2d}}\frac{d-1}{d}\left(B_{rr}^{n} - B_{\theta\theta}^{n}\right) w_3 \diff r, \\
    \int_\Omega S_{\theta\theta}^nw_4 \diff r &= \int_\Omega \kappa\left(\tilde{\phi}^{n-1}\right) \left(\left(B_{rr}^n\right)^{1/2} \left(B_{\theta\theta}^n\right)^{(d-1)/2} -1\right)w_4 \diff r \\
    &+ \int_\Omega G\left(\tilde{\phi}^{n-1}\right) \left(B_{rr}^{n}\left(B_{\theta\theta}^{n}\right)^{d-1}\right)^{-\frac{2+d}{2d}} \frac{1}{d}\left(-B_{rr}^{n} + B_{\theta\theta}^{n}\right) w_4 \diff r, \\
    \text{with } \kappa\left(\tilde{\phi}^{n-1}\right) &= \kappa\sub{in}\frac{1+\tilde{\phi}^{n-1}}{2}  + \kappa\sub{out} \frac{1-\tilde{\phi}^{n-1}}{2}, G\left(\tilde{\phi}^{n-1}\right) = G\sub{in}\frac{1+\tilde{\phi}^{n-1}}{2}  + G\sub{out} \frac{1-\tilde{\phi}^{n-1}}{2}.
\end{align*}

Again, note that the stress is coupled to the strain by $B_{rr}^n$ and $B_{\theta\theta}^n$ and to the Cahn-Hilliard equation by $\tilde{\phi}^{n-1}$. Nonlinear terms appear in both the isotropic and deviatoric parts of the stress equations. These are linearized using first-order Taylor expansions, which are not written out here.

In order to accurately resolve the forces at the interface we utilize an adaptive grid. The interfacial grid is refined heuristically, determined by the phase-field value. In each time step, the position of the interface is determined, i.e. where $\phi = 0$. Within a fixed interval around the interface, the grid is refined according to the smaller grid width $h_\text{interface}$. In a larger interval around the interface an intermediate grid width $h_\text{transition}$ is applied. Elsewhere, the grid width is set to $h_\text{bulk}$ with $h_\text{interface} \le h_\text{transition} \le h_\text{bulk}$.

\section*{Validation of the Finite Difference Method}

In the main text, we validated the phase-field model for a 1D radially symmetric scenario, enabling direct comparison with \cite{wei2023eulerian}. Here, we extend this validation to the model and numerical method in two dimensions. In the absence of a reference solution, we rely on comparison with the previously validated 1D radially symmetric method. However, as seen in the main text, the 2D solutions yield tumor shapes that deviate from a circular form. Consequently, we restrict our comparison to selected high relaxation rates ($\beta\sub{in} = \beta\sub{out} = 400$) and early times (time interval $[0,20]$), before protrusion formation occurs. 
Accordingly, we compare results from the 2D Finite Difference Method (FDM) and the 1D Finite Element Method (FEM) in the following. To ensure fair comparison with the 2D results at larger values of $\epsilon$, radially symmetric terms are included in the 1D Cahn-Hilliard equation.

\begin{figure}
\centering
\includegraphics[width=1\linewidth]{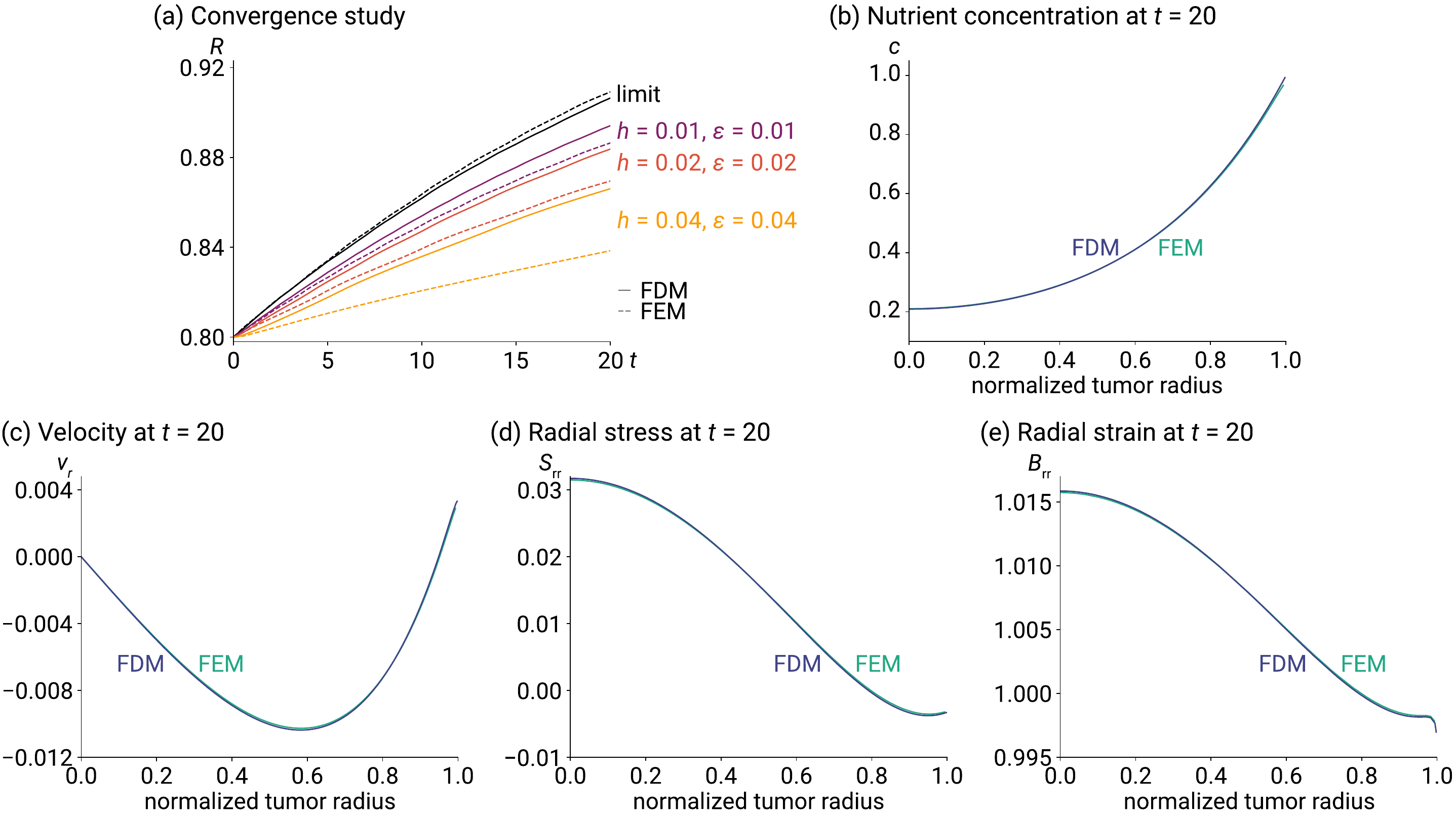} 

\caption{Comparison of the 2D FDM and the FEM. (a) Convergence of the radius over time in the grid width $h$ and the interface thickness $\epsilon$. The experimentally determined order of convergence of the FDM is 0.85 and of the FEM 0.82. From these values we extrapolate the limit curves (black) which show good agreement. (b)--(e) Comparison of the results at $t=20$ for the finest resolution $h=0.01, \epsilon = 0.01$. Simulation parameters: domain FEM: $R_\Omega=2, R_0=0.8$, domain FDM: $L_x=L_y=4$; $\zeta=6.25, M=0.01, \kappa\sub{in} = 2.0, G\sub{in} = 1.0, \kappa\sub{out} = 0, G\sub{out} = 0, \beta\sub{in} = \beta\sub{out} = 400, \lambda\sub{p}=0.2, \lambda\sub{a}=0.1, L=0.3$.}
\label{fig:Validation_FDM}
\end{figure}

As shown in Fig. \ref{fig:Validation_FDM}, the errors in the solutions are relatively small and seem attributable to the different discretization strategies employed by the two methods. For the best refinement, we find almost perfect agreement of nutrient concentration, velocity, strain and stress (Fig.~\ref{fig:Validation_FDM}b-e).
Although the evolution of tumor size (Fig.~\ref{fig:Validation_FDM}a) displays larger discrepancies across different resolutions, the data suggests convergence as $\epsilon,h\rightarrow 0$. To rigorously assess this behavior, we calculated the numerical order of convergence at the final time, obtaining a value of 0.85 for the FDM and 0.82 for the FEM.
Using these values, we computed the numerical convergence limit through Richardson extrapolation, as shown by the black curves in Fig. \ref{fig:Validation_FDM} (a). At final time, the FDM limit radius is 0.908 and the FEM limit radius is 0.902, indicating that the error is below 1 \%.

\section*{Convergence with respect to the diffuse interface parameter} \label{sec: channels and eps convergence in 2D}

To assess convergence with respect to the diffuse interface parameter for \textit{non-spherical} shapes, we compare tumor morphologies for various $\epsilon$. 
As $\epsilon$ has a significant influence on the onset of finger formation, we perform the comparison for time points at which tumors have approximately the same size.

\begin{figure} 
    \centering
    \includegraphics[width=0.7\textwidth]{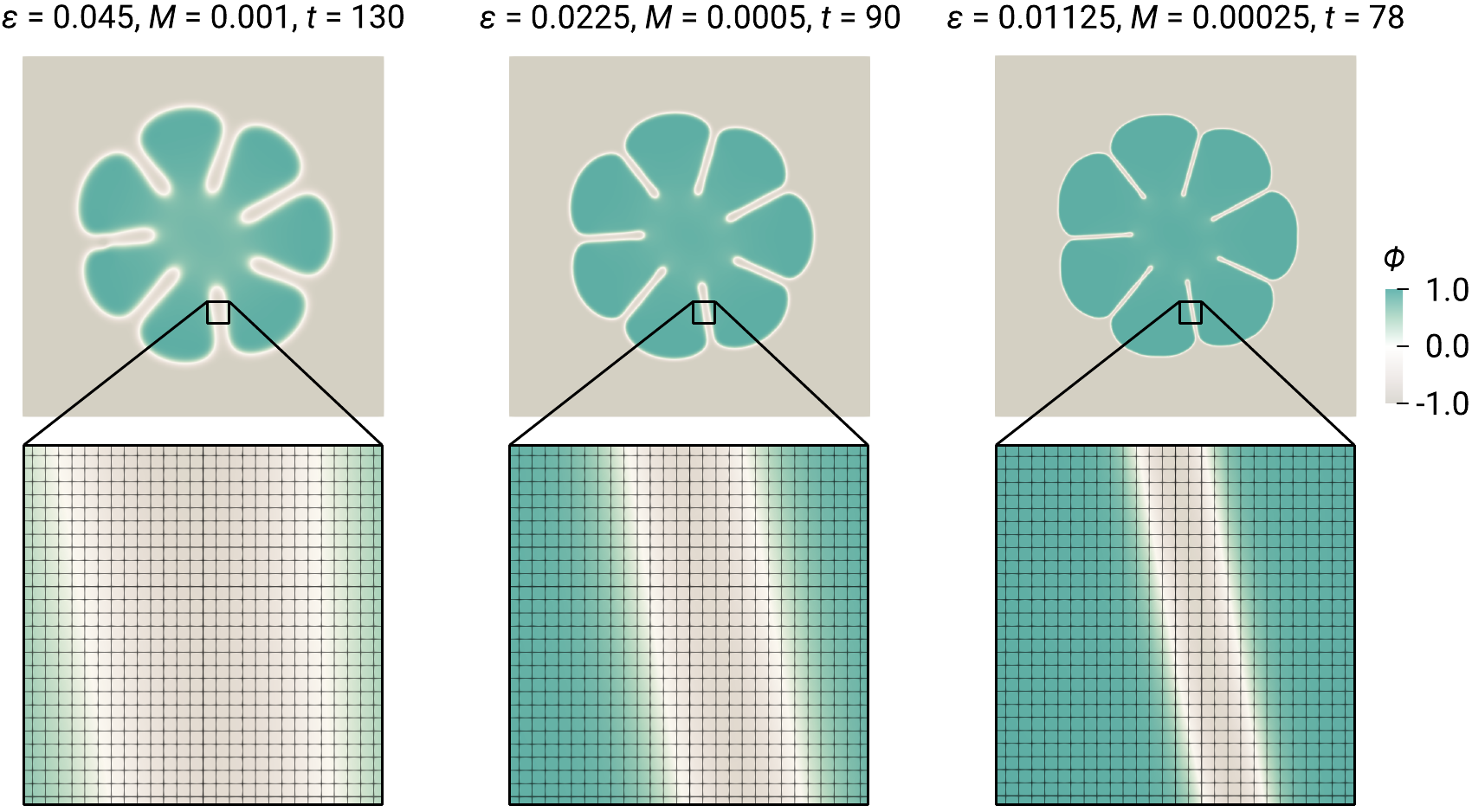}
    \caption{Comparison of tumor morphologies across different values of diffuse interface width ($\epsilon$) demonstrates overall convergence as $\varepsilon\rightarrow 0$ towards a similar final morphology. The width of the channels forming between the outgrowing protrusions is found to scale linearly with $\epsilon$ (see bottom panels). 
    The three columns correspond to different time points at which the tumors cover approximately the same area (error $< 0.5\%$). Further simulation parameters: $L_x=L_y=5.2, h=0.01, R_0=0.8, \zeta=6.25, \kappa\sub{in} = 2, G\sub{in} = 1, \kappa\sub{out} = G\sub{out} = 0, \beta\sub{in} = \beta\sub{out}=0.3, \lambda\sub{p} = 0.2, \lambda\sub{a} = 0.1, L= 0.3.$}
    \label{fig:channel_width}
\end{figure}

The results in Fig.~\ref{fig:channel_width} suggest convergence, as $\epsilon\rightarrow 0$, towards a morphology which is quite similar to the three shapes observed in our test cases. This agreement validates the use of the fixed value $\epsilon=0.045$ throughout this work.

The main difference between shapes seems to be the width of the channels between the outgrowing protrusions. 
Notably, this width is found to scale linearly with $\epsilon$ (see bottom panels).
The phase-field profile in these channels approaches a minimum value of $-0.8$, indicating that the two neighboring diffuse interfaces 'feel' each other through the soft elastic material in between them. 
This presence of elastic material trapped between the two interfaces provides an effective repulsion mechanism: 
Any further narrowing of the channel would require to drain out the trapped material, which would, in turn, induce substantial shear stress.
Having $G\sub{in}>0$ opposes this shear stress wherever $\phi>-1$ and keeps the interfaces separated at a fixed distance. 

\section*{Role of surrounding tissue in tumor surface instabilities}
\label{sec:instabilities_appendix}

As discussed in the main text, tumor growth exhibits different behaviors under confined versus unconfined conditions. In particular, we observed that, in the absence of shear elasticity and apoptosis, the tumor shows surface instabilities when host tissue is present, unlike the case without host tissue, where growth is isotropic. Here, we aim to investigate the underlying mechanism that leads to protrusion formation when external material is present.

\begin{figure}[t] 
    \centering
    \includegraphics[width=0.9\textwidth]{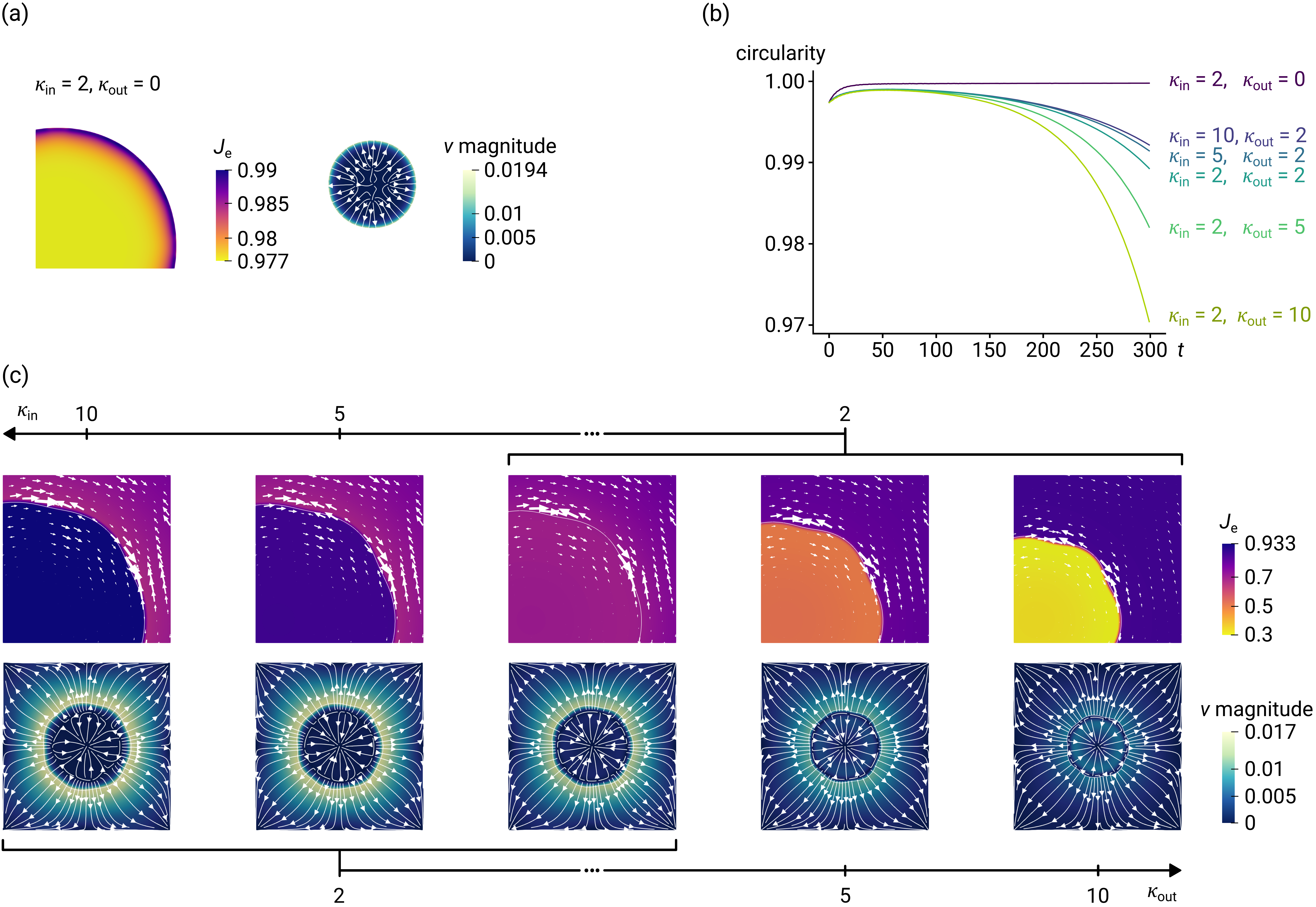}
    \caption{Tumor growth patterns at $t=240$ without shear elasticity and apoptosis ($G\sub{in}=G\sub{out}=0, \lambda\sub{a}=0$). \textbf{(a)} Unconfined case ($\kappa\sub{out}$=0); panels show $J_e$ as well as the direction and magnitude of velocity $\textbf{v}$. \textbf{(b)} Circularity -- i.e., deviation of the shape from a circle -- computed as $\frac{2\sqrt{\pi \text{ area}}}{\text{perimeter}}$, for different combinations of $\kappa\sub{in}, \kappa\sub{out}$ shown in (a) and (b). \textbf{(c)} confined case, varying $\kappa\sub{in}$ and $\kappa\sub{out}$. Top panels show  $J_e$ and the tangential (azimuthal) component of the velocity $\textbf{v}$; bottom panels show the direction and magnitude of the full velocity. Further simulation parameters: $L_x=L_y=16, h=0.04, R_0=0.8, \epsilon=0.045, \zeta=6.25, M=0.001, \beta\sub{in} = \beta\sub{out}=0.3, \lambda\sub{p} = 0.06, L= 0.3$. }
    \label{fig:instabilities_constrained_appendix}
\end{figure}

In the case without surrounding tissue (\figref{fig:instabilities_constrained_appendix}a), the tumor grows isotropically when $\kappa\sub{out} = \lambda\sub{a} = 0$, as indicated by the constant value of tumor circularity (\figref{fig:instabilities_constrained_appendix}b). The velocity vectors point radially outward, reflecting radially symmetric expansion. In contrast, when external material is present (\figref{fig:instabilities_constrained_appendix}c, $\kappa_\text{out} > 0$), instabilities appear at the tumor surface, manifested as indentations or protrusions, with flow directed inward toward the tumor center.

To test whether these inward flows are the origin of the observed instabilities, we first varied the tumor’s compressibility $\kappa\sub{in} = 2, 5, 10$, while keeping $\kappa\sub{out} = 2$. For larger $\kappa\sub{in}$ we find that inward flows significantly reduce in magnitude. However, no significant effect on protrusion formation was observed: although the tumor is less compressed (higher $J_e$) for larger $\kappa\sub{in}$, the shape instabilities remain similar. This is also reflected in the circularity curves (\figref{fig:instabilities_constrained_appendix}b) and indicates that inward flows are not the primary driver of the shape instability.

Next, we varied the compressibility of the surrounding material ($\kappa\sub{out} = 2, 5, 10$) while keeping $\kappa\sub{in} = 2$. Increasing $\kappa\sub{out}$ leads to stronger compression of the tumor (smaller $J_e$), as the stiffer host tissue restricts the tumor’s ability to  expand. Compared to the variation of $\kappa\sub{in}$, the change in $\kappa\sub{out}$ has a more noticeable influence on the tumor shape, with the circularity gradually decreasing as $\kappa_\text{out}$ increases (\figref{fig:instabilities_constrained_appendix}b).

These results indicate that the elasticity and compressibility of the \textit{surrounding} tissue are the primary drivers of the observed protrusion formation (\figref{fig:instabilities_constrained_appendix}c).
The mechanism behind this phenomenon  is elucidated by examining the tangential velocity shown in the top panels of \figref{fig:instabilities_constrained_appendix}c. We find that near the protrusions at the tumor boundary, the flow is directed toward the indentations, while in the interior flow is oriented oppositely. 
Hence, protrusion formation leads to a circular motion: outgrowing protrusions push the surrounding host tissue sideways, and the host tissue's bulk elastic resistance, in turn, exerts pressure on the tumor boundary adjacent to the protrusion, thereby contributing to tumor invagination there. Together with differential growth, this mechanism seems to amplify formation of protrusions and invaginations over time. 

\section*{Influence of tissue fluidization at the tumor boundary}
\label{sec:fluidization_at_boundary}

\begin{figure}
    \centering
    \includegraphics[width=0.8\linewidth]{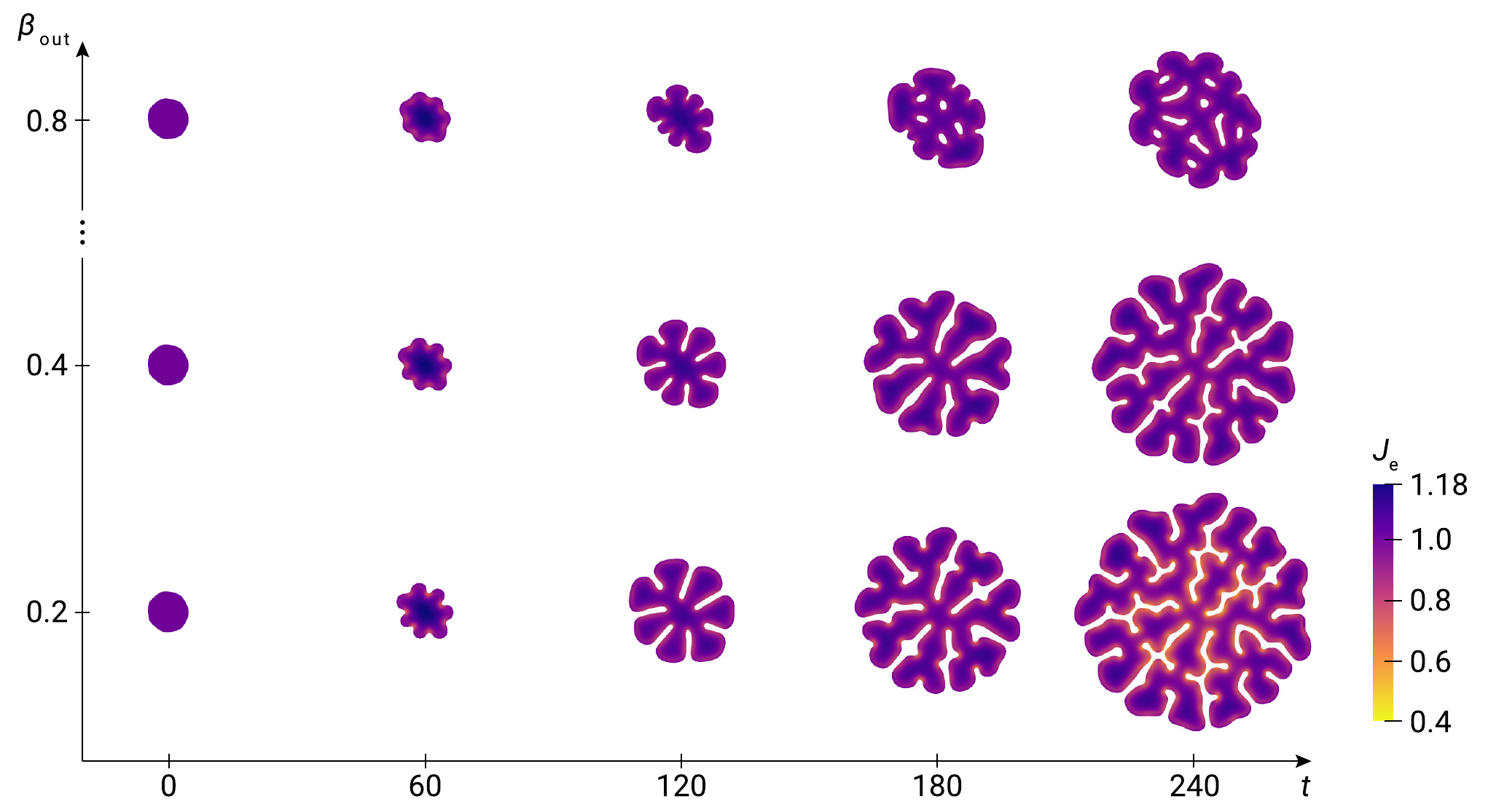}
    \caption{Influence of surface fluidization: comparison of tumor evolution depending on relaxation along the tumor periphery ($\beta\sub{out} = 0.8, 0.4, 0.2$ from top to bottom). Tumor shapes and $J_e$ at $t = 0, 60, 120, 180, 240$. Further parameters: $L_x=L_y=12, h = 0.04, R_0=0.8, \epsilon = 0.045, \zeta=6.25,M = 0.001,  \kappa\sub{in} = 2, G\sub{in} = 1, \kappa\sub{out} = G\sub{out} = 0, \beta\sub{in} = 0.1, \lambda\sub{p} = 0.2, \lambda\sub{a} = 0.1, L= 0.3$. }
\label{fig:study_beta_out_unconstrained_perturbed}
\end{figure}

Growing tumors can secrete enzymes which contribute to  degradation of biological material, such as ECM and cell-cell connections. 
Since these  enzymes preferentially work at the tumor periphery, they give rise to an effective fluidization along the tumor boundary. 
We model this behavior by varying $\beta$ along the tumor periphery. This can be  achieved in our phase-field model by varying $\beta\sub{out}>\beta\sub{in}$, leading to a gradual increase of $\beta$ over the diffuse interface. 
The results in Fig.~\ref{fig:study_beta_out_unconstrained_perturbed} show the influence of surface fluidization. 
Consistent with its effect in the interior, fluidization of the tumor periphery slows down protrusion formation. 

In addition, increased surface fluidity accelerates the onset of topological changes.
This behavior is driven by the reduction in shear resistance; the rapid dissipation of shear stresses between protrusions allows the soft material in between to flow out more efficiently. This accelerated stress relaxation facilitates the coalescence of neighboring structures, thereby promoting earlier topological transitions."

\section*{Influence of surface tension}
\label{surface tension}

In the current model, it is possible to specify additional forces acting on the tumor. 
One such force is surface tension, which models cell-cell adhesive forces at the tumor-host interface, e.g. \cite{byrne1996,byrne1997,cristini2003,Pham2018}. 
As a tumor grows, it exerts mechanical forces on its surrounding tissue, leading to changes in tissue stiffness and deformation \cite{angeli2025}. These changes can influence the behavior of the tumor cells and their interaction with the surrounding environment. Additionally, the ability of tumor cells to invade and migrate into surrounding tissues is influenced by mechanical properties such as tissue stiffness and the presence of physical barriers. These properties can affect the surface tension-like forces that govern cell-cell and cell-matrix interactions at the tumor periphery. 
To include it, we introduce a surface tension force 
\begin{equation*}
    F_\sigma = \sigma \mu \nabla \phi,
\end{equation*}
where $\sigma$ is a scaling factor, and add it to the force balance,
\begin{equation}
    \zeta\mathbf{v} = \nabla \cdot \mathbf{S} + F_\sigma.
\end{equation}

\begin{figure}
    \centering
    \includegraphics[width=0.6\linewidth]{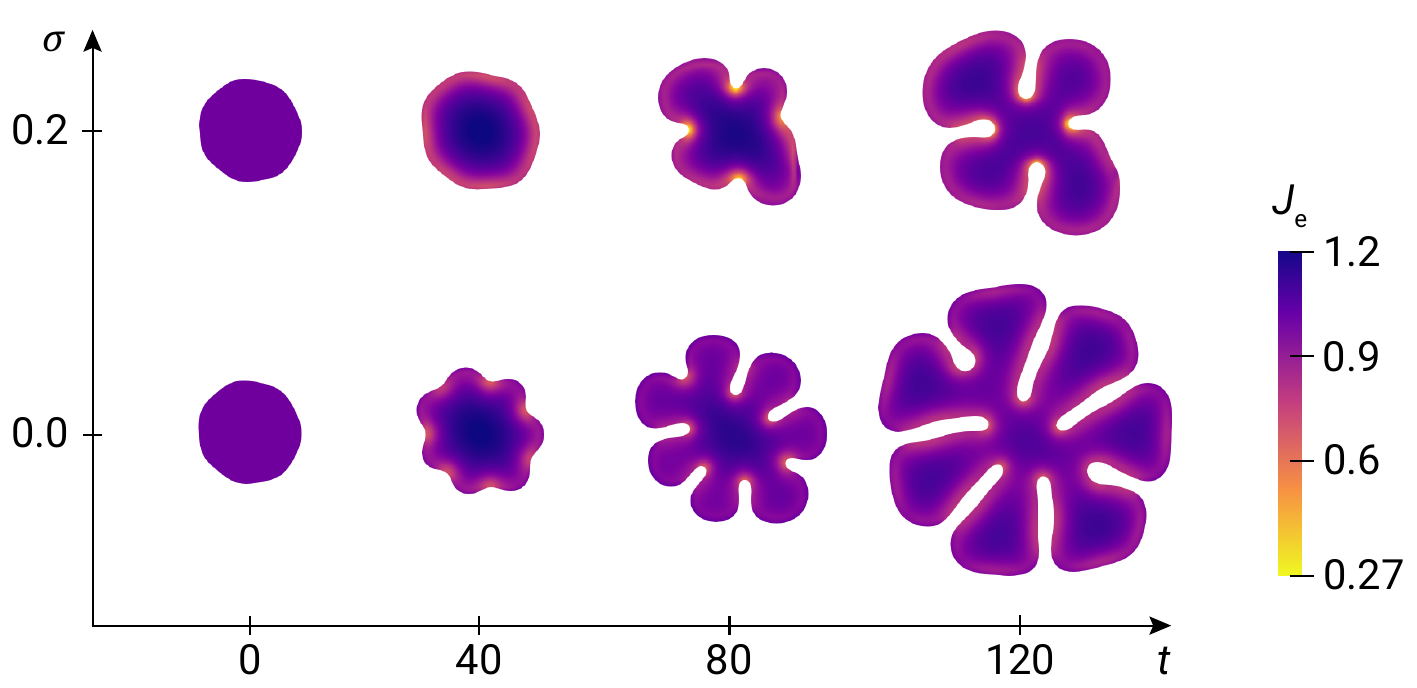}
    \caption{Influence of surface tension (top: $\sigma = 0.2$, bottom: $\sigma = 0$). Tumor shapes and $J_e$ at $t = 0, 40, 80, 120$. Further parameters: $L_x=L_y= 8, h=0.04, R_0=0.8, \epsilon=0.045, \zeta=6.25, M=0.001, \kappa\sub{in}=2, G\sub{in}=1, \kappa\sub{out}=G\sub{out}=0, \beta\sub{in} = \beta\sub{out} = 0.1, \lambda\sub{p} = 0.2, \lambda\sub{a} = 0.1, L= 0.3$.}
\label{fig:surface_tension}
\end{figure}

We find that the presence of surface tension, on one hand, slows down the growth of the tumor tissue. On the other hand, it influences the emerging outgrowth patterns. As shown in Fig.~\ref{fig:surface_tension}, a tumor with initial modes 2 and 7 develops seven protrusions when surface tension is turned off, whereas only four protrusions emerge when surface tension is present. The surface tension force tends to push the tissue back into a compact shape with minimal surface area, which is why surface tension reduces the number of protrusions in the growth pattern.

\section*{Supplemental Movies}

\textbf{Movie S1:} Simulation of a growing spherical unconfined tumor. The 1D radially symmetric solution is shown using a 2D circular representation. The panels (from left to right) show the elastic volume variation $J_e$, the velocity $v_r$ and the nutrient concentration $c$. Because the overall proliferation exceeds apoptosis the tumor grows over time. The increase in tumor size progressively reduces the nutrient concentration in the center, slowing down overall growth until the tumor radius converges to a stationary value. In the stationary state, a proliferating rim persists at the tumor boundary, where proliferating material is transported inward toward the tumor center to compensate for local apoptosis. Simulation parameters: $R_\Omega=3, R_0=1.3, \varepsilon=0.0025, \zeta=6.25, M=1.6$,  $\kappa_\text{in}=2, G_\text{in}=1, \kappa_\text{out}=G_\text{out}=0, \beta=0.3$, $\lambda\sub{p}=0.2, \lambda\sub{a}=0.1, L=0.3$. \\

\noindent\textbf{Movie S2:} Evolution of a compressible unconfined 2D tumor with perturbed initial condition (modes 2 and 7). Small values of $J_e$ at the boundaries indicate compression ($J_e<1$) and high density ($\rho=\rho_0/J_e$). During growth, the tumor develops pronounced protrusions resulting from fingering and buckling instabilities. Simulation parameters: $L_x=L_y=16, h=0.04, R_0=0.8, \epsilon=0.045$, $\zeta=6.25$, $M=0.001$, $\kappa_\text{in}=2, G_\text{in}=1, \kappa_\text{out}=G_\text{out}=0, \beta_\text{in}=\beta_\text{out}=0.3,$ $\lambda\sub{p} = 0.2$, $\lambda\sub{a} = 0.1$, $L= 0.3$. \\

\noindent\textbf{Movie S3:} Evolution of a highly fluidized compressible unconfined 2D tumor with perturbed initial condition (modes 2 and 7) showing early changes in topology. Simulation parameters: $L_x=L_y=6.4, h=0.04, R_0=0.8, \epsilon=0.045$, $\zeta=6.25$, $M=0.001$, $\kappa_\text{in}=2, G_\text{in}=1, \kappa_\text{out}=G_\text{out}=0, \beta_\text{in}=\beta_\text{out}=5.0,$ $\lambda\sub{p} = 0.2$, $\lambda\sub{a} = 0.1$, $L= 0.3$. \\

\noindent\textbf{Movie S4:} Simulation of a 3D tumor exhibiting fingering and buckling behavior. Tumor contour in 3D shown, initially perturbed with a sinusoidal perturbation with frequency parameter $8$ along both polar angles. Simulation parameters: $L_x=L_y=L_z=7, h=0.05, R_0=1.3, \epsilon=0.05, \zeta=6.25, M=0.001, \kappa\sub{in} = 2, G\sub{in} = 1, \kappa\sub{out} = G\sub{out} = 0, \beta\sub{in} = \beta\sub{out}=0.3, \lambda\sub{p} = 0.2, \lambda\sub{a} = 0.1, L= 0.3.$  

\end{document}